%% file: main.tex
\documentclass[final,5p]{elsarticle}
\pdfoutput=1 
\usepackage[switch]{lineno}
\usepackage{hyperref}
\hypersetup{hidelinks}
\usepackage{amsmath}
\usepackage{amssymb}
\usepackage{gensymb}
\usepackage{amstext}
\usepackage{mathtools}
\usepackage{esint}
\usepackage{caption}
\usepackage{wrapfig}
\usepackage{float}
\usepackage{dblfloatfix}
\usepackage[symbol]{footmisc}
\journal{Nuclear Instrumentation and Methods}
\usepackage{geometry}
\usepackage{upgreek}
\usepackage{epsfig}
\usepackage{subcaption}
\newcommand{\microns}{\(\upmu\)m}
\usepackage{array}
\newcolumntype{P}[1]{>{\centering\arraybackslash}p{#1}}

\usepackage{titlesec}

\titleformat{\chapter}
  {\normalfont\fontsize{16}{19}\bfseries}
  {\thechapter}
  {1em}
  {}

\titleformat{\section}
  {\normalfont\fontsize{12}{17}\bfseries}
  {\thesection}
  {1em}
  {}

\titleformat{\subsection}
  {\normalfont\fontsize{11}{16}\bfseries\slshape}
  {\thesubsection}
  {1em}
  {}

\titleformat{\subsubsection}
  {\normalfont\fontsize{10}{12}\bfseries\slshape}
  {\thesubsubsection}
  {1em}
  {}

\begin{document}


\begin{frontmatter}

\title{Design and production of the high voltage electrode grids and electron extraction region for the LZ dual-phase xenon time projection chamber}

\input{LZAuthorList.tex}



\begin{abstract}
The dual-phase xenon time projection chamber (TPC) is a powerful tool for direct-detection experiments searching for WIMP dark matter, other dark matter models, and neutrinoless double-beta decay. Successful operation of such a TPC is critically dependent on the ability to hold high electric fields in the bulk liquid, across the liquid surface, and in the gas. Careful design and construction of the electrodes used to establish these fields is therefore required. We present the design and production of the LUX-ZEPLIN (LZ) experiment's high-voltage electrodes, a set of four woven mesh wire grids. Grid design drivers are discussed, with emphasis placed on design of the electron extraction region. We follow this with a description of the grid production process and a discussion of steps taken to validate the LZ grids prior to integration into the TPC.
\end{abstract}

\begin{keyword}
Xenon \sep TPC \sep HV \sep Electrode \sep Noble Liquid
\end{keyword}

\end{frontmatter}

\setcounter{footnote}{0}

\sloppy
\input{0_Introduction/Introduction.tex}

\input{2_GridERDesign/GridERDesign.tex}

\input{3_GridProduction/GridProduction.tex}

\input{4_GridERValidation/GridERValidation.tex}

\input{7_Conclusions/Conclusions.tex}
\input{8_Acknowledgements/Acknowledgements}

\bibliography{GridProduction}{}

\end{document}

%% file: LZAuthorList.tex
\author[1,2]{R.~Linehan\corref{cor1}}
\author[6]{R.L.~Mannino\corref{cor1}}
\cortext[cor1]{Corresponding authors: rlinehan@stanford.edu and mannino2@wisc.edu}
\author[1,2]{A.~Fan}
\author[1,2]{C.M.~Ignarra}
\author[1]{S.~Luitz}
\author[1]{K.~Skarpaas}
\author[1,2]{T.A.~Shutt}

\author[1,2]{D.S.~Akerib}
\author[6]{S.K.~Alsum}
\author[1,2]{T.J.~Anderson}
\author[8]{H.M.~Ara\'{u}jo}
\author[3]{M.~Arthurs}
\author[1]{H.~Auyeung}
\author[8]{A.J.~Bailey}
\author[1,2]{T.P.~Biesiadzinski}
\author[1]{M.~Breidenbach}
\author[6]{J.J.~Cherwinka}
\author[1]{R.A.~Conley}
\author[34,10]{J.~Genovesi} 
\author[13]{M.G.D.~Gilchriese}
\author[1,2]{A.~Glaenzer}
\author[1]{T.G.~Gonda}
\author[13]{K.~Hanzel}
\author[13]{M.D.~Hoff}

\author[1,2]{W.~Ji}
\author[35,12]{A.C.~Kaboth}
\author[13]{S.~Kravitz}
\author[1]{N.R.~Kurita}
\author[13]{A.R.~Lambert}
\author[13]{K.T.~Lesko}

\author[3]{W.~Lorenzon}

\author[12]{P.A.~Majewski}

\author[1,2]{E.H.~Miller}
\author[1,2]{M.E.~Monzani}
\author[6]{K.J.~Palladino}
\author[1]{B.N.~Ratcliff}
\author[13]{J.S.~Saba}
\author[35]{D.~Santone}
\author[1]{G.W.~Shutt}



\author[1,2]{K.~Stifter}
\author[34]{M.~Szydagis}
\author[8]{A. Tom\'{a}s}
\author[1]{J.~Va'vra}
\author[13]{W.L.~Waldron}
\author[39]{R.C.~Webb}
\author[1,2]{R.G.~White}
\author[1,2]{T.J.~Whitis}
\author[13]{K.~Wilson}
\author[1]{W.J.~Wisniewski}


\address[1]{SLAC National Accelerator Laboratory, Menlo Park, CA 94025-7015, USA}

\address[2]{Kavli Institute for Particle Astrophysics and Cosmology, Stanford University, Stanford, CA  94305-4085 USA}

\address[6]{University of Wisconsin-Madison, Department of Physics, Madison, WI 53706-1390, USA}

\address[8]{Imperial College London, Physics Department, Blackett Laboratory, London SW7 2AZ, UK}

\address[3]{University of Michigan, Randall Laboratory of Physics, Ann Arbor, MI 48109-1040, USA}

\address[34]{University at Albany (SUNY), Department of Physics, Albany, NY 12222-1000, USA}

\address[13]{Lawrence Berkeley National Laboratory (LBNL), Berkeley, CA 94720-8099, USA}

\address[35]{Royal Holloway, University of London, Department of Physics, Egham, TW20 0EX, UK}

\address[12]{STFC Rutherford Appleton Laboratory (RAL), Didcot, OX11 0QX, UK}

\address[10]{South Dakota School of Mines and Technology, Rapid City, SD 57701-3901, USA}

\address[39]{Texas A\&M University, Department of Physics and Astronomy, College Station, TX 77843-4242, USA}

%% file: 0_Introduction/Introduction.tex
\section{Introduction}
\label{sec:intro}

\begin{figure*}[htbp]
\centering
\includegraphics[width=13cm]{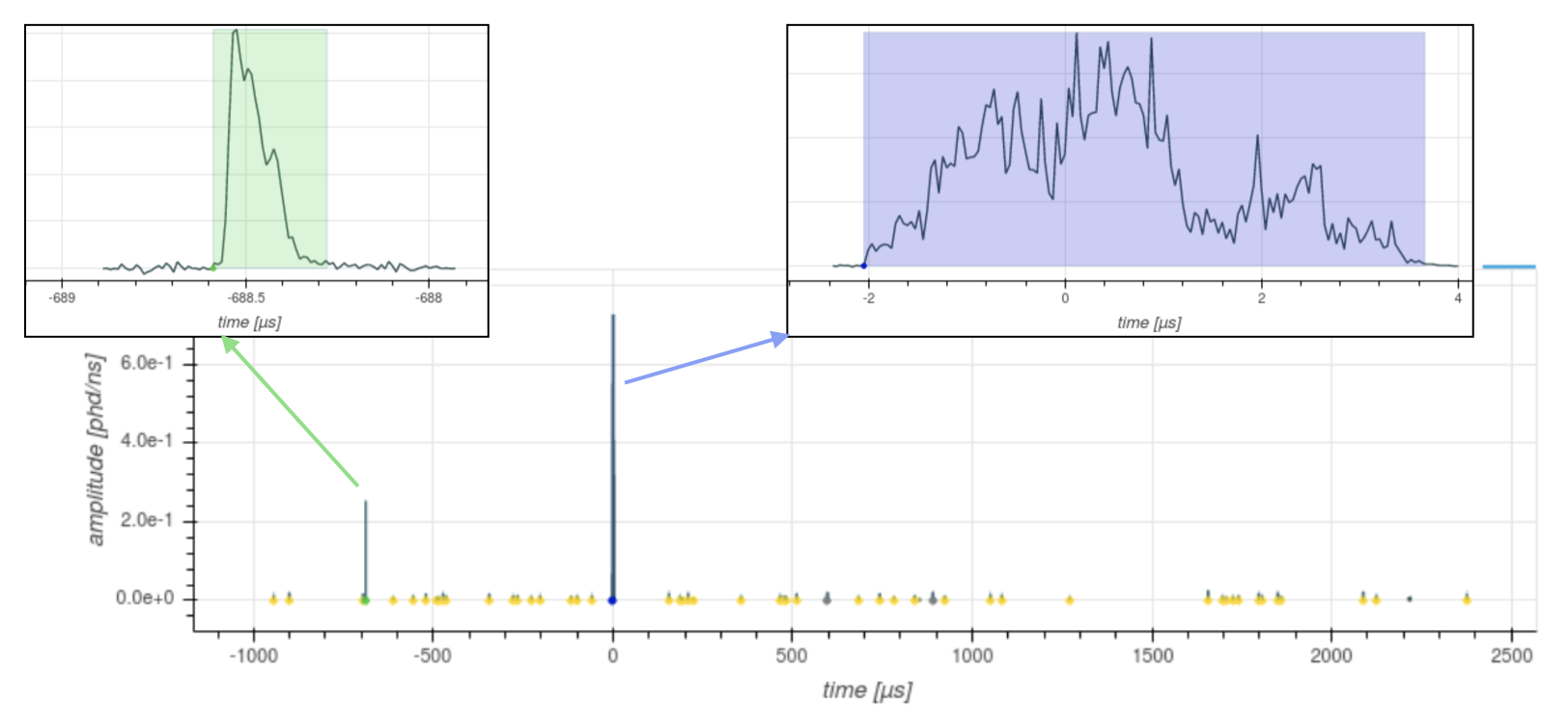}
\caption{
A waveform of a sample low-energy scatter event simulated in the LZ detector geometry\cite{BACCARAT}. This waveform is produced by summing the light signals observed in every photomultiplier tube (PMT). The S1 (green) and S2 (blue) pulses are separated by 700 $\upmu$s, which together with the electron drift velocity in LXe indicates the depth of the interaction in the detector.}
\label{S1S2}
\end{figure*} 

The LUX-ZEPLIN (LZ) experiment is a low-background rare event search detector located 4850 feet below the surface at the Sanford Underground Research Facility (SURF) in Lead, South Dakota, USA~\cite{LZDetector}. Its primary goal is to search for scatters of WIMP dark matter \cite{rpp2019} on xenon nuclei. However, it will also conduct other rare event searches, including those for coherent neutrino-nucleus scatters (CE$\nu$NS) of solar $^{8}$B neutrinos and supernova neutrinos on Xe \cite{B8}, axion-like particle (ALP) interactions, and neutrinoless double beta (0$\nu\beta\beta$) decay from $^{136}$Xe \cite{DoubleBetaDecaySensitivity}. The variety of rare physics searches requires LZ to detect and reconstruct interactions across a wide range of recoil energies, from a few keV for scatters of dark matter candidates with masses of approximately 1~GeV/$c^{2}$, to 2.458~MeV for 0$\nu\beta\beta$ decay from \(^{136}\)Xe.

\begin{figure*}[htb]
\centering
\includegraphics[width=6in]{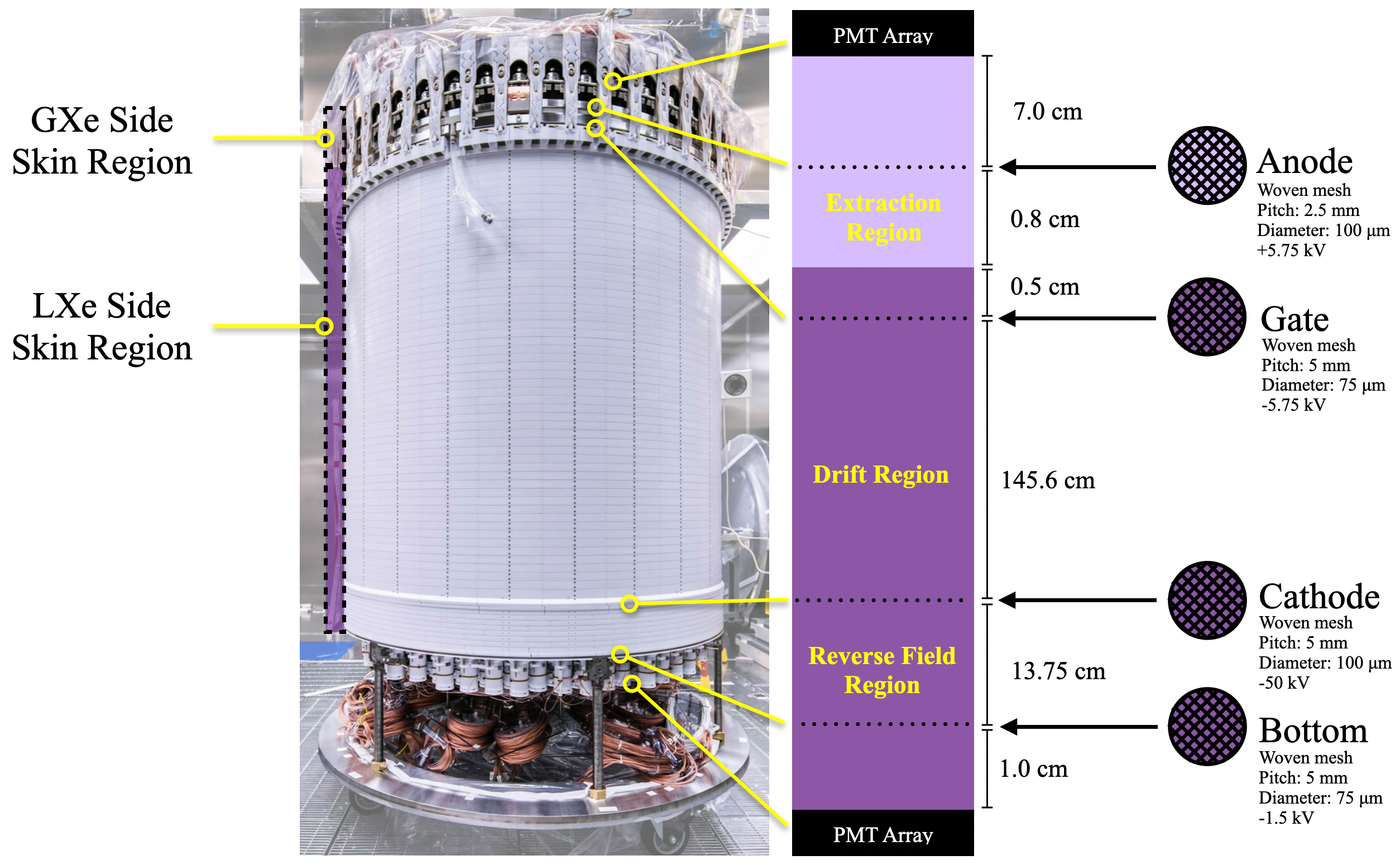}
\caption{The HV electrode configuration for the LZ TPC, shown next to a photograph of the fully assembled TPC. In the diagrams, light purple implies gaseous xenon, and dark purple implies liquid xenon. The electroluminescence field (10~kV/cm) is located between the liquid surface and the anode, and the extraction field (5~kV/cm) is located in between the gate and the liquid surface. The locations and sizes of the wires in the diagram are not to scale.}
\label{TPCDiagram}
\end{figure*}

LZ's inner detector is a dual-phase xenon time projection chamber (TPC) with a 7 tonne active (5.6 tonne fiducial) target mass. Energy depositions in the bulk liquid xenon (LXe) create two light signals detected by arrays of photomultiplier tubes (PMTs) at the top and bottom of the TPC. An example of a simulated event containing these two signals is shown in Figure \ref{S1S2}. The first signal (S1) is a prompt, sharply-peaked flash of scintillation light observed within approximately 100 ns of an interaction. The second signal (S2) is created once free electrons from ionization following the initial interaction drift upwards through the bulk xenon. This drift is accomplished by an electric field created between a cathode electrode grid and gate electrode grid, which straddle the bulk of the liquid xenon (Figure~\ref{TPCDiagram}). Upon reaching the liquid surface, these electrons are extracted into a thin gas-phase xenon region. There, they produce electroluminescence light in a strong electric field created between the gate grid (below) and an anode grid above. The width of the S2 pulse is partially determined by the time that each electron spends drifting in the gas, and is partially determined by the vertical spatial extent of the electron cloud when it reaches the surface. For single electrons (SEs), the pulse width is expected to be approximately 1 $\upmu$s. The position of the interaction in the plane perpendicular to the TPC axis (the \textit{xy} plane) can be reconstructed to within about 0.5~cm\footnote{This estimate varies with S2 size, and is quoted for the WIMP search S2 threshold size.} using the pattern of detected electroluminescence (S2) photons in the top PMT array. The time difference between the S1 and S2 can be as large as 800 $\upmu$s, and can be used to find the depth of the interaction below the liquid surface. By combining \textit{xy} and depth information, three-dimensional event position reconstruction is achievable. This allows discrimination of potential dark matter interactions, which should occur uniformly throughout the detector, from peripheral interactions caused by radioactivity in the cavern walls, in the detector materials, and on detector component surfaces. An instrumented ``xenon skin" region just outside of the TPC but within the inner cryostat vessel (ICV) provides additional shielding from such radiation, and enhances the ability to reject multiply-scattering gamma-rays from ambient radioactivity.

LZ critically depends on its electrode grids to create strong, uniform electric fields for drifting free electrons and creating the S2 light. These grids span the 1.456~m diameter of the TPC, and in doing so promote drift and extraction field uniformity. However, this full \textit{xy} coverage comes at a cost -- the grids partially obscure S1 and S2 light, and may also create field-induced and radiogenic backgrounds that are challenging to reject (Section \ref{subsec:EngineeringConstraints}). To ensure that the grids could successfully create strong enough drift and extraction fields without substantially reducing WIMP search sensitivity in these ways, a dedicated grid design and production program was conducted between 2014 and 2019.

In Section \ref{sec:gridERDesign} of this paper we motivate and discuss the design of the LZ high-voltage electrodes. We also describe the design of the extraction region, which is responsible for the creation of the S2 signal. We note however that this work does not cover the design of the complex cathode HV connection, described in Ref~\cite{LZDetector}. In Section \ref{sec:gridProduction}, we discuss the environment and process used in electrode fabrication. We finish with Section \ref{sec:validation}, where we discuss the steps taken to validate the grids and extraction region ahead of installation in the TPC.

%% file: 2_GridERDesign/GridERDesign.tex
\section{Grid and Extraction Region Design}
\label{sec:gridERDesign}

\subsection{Drivers of Electric Field Design}
\label{subsec:EFieldDesignDrivers}

Optimal performance of a dual-phase xenon TPC requires electric fields that are strong enough to remove free electrons from an interaction site, drift them through the liquid, extract them into the gas, and create the electroluminescence signal. At the same time, it is important to maximize field uniformity and minimize stray fields that could induce electrical breakdown.

Efficient discrimination between nuclear recoils and electron recoils using a charge-to-light ratio favors maintaining a drift field between 240 and 290~V/cm~\cite{LUXDiscrimination}. Lower drift fields reduce discrimination between the two types of recoils. Lower fields also decrease the electron drift velocity, which falls off more quickly with decreasing field below about 100 V/cm~\cite{EXODriftVelocity}. This increases the average time between an S1 and an S2, resulting in an undesirable increase in the rate at which uncorrelated interactions pile up into a single drift window. A lower drift velocity also increases the time $t_{d}$ over which an electron cloud will drift, resulting in a diffusion-induced S2 pulse width increase that goes as $\sqrt{t_{d}}$. These changes can make S2 pulse identification and S1-S2 association more challenging, which can reduce WIMP scatter detection efficiency. All of these factors motivate maintaining a strong drift electric field.

\begin{figure*}[t!]
\centering
\includegraphics[width=6in]{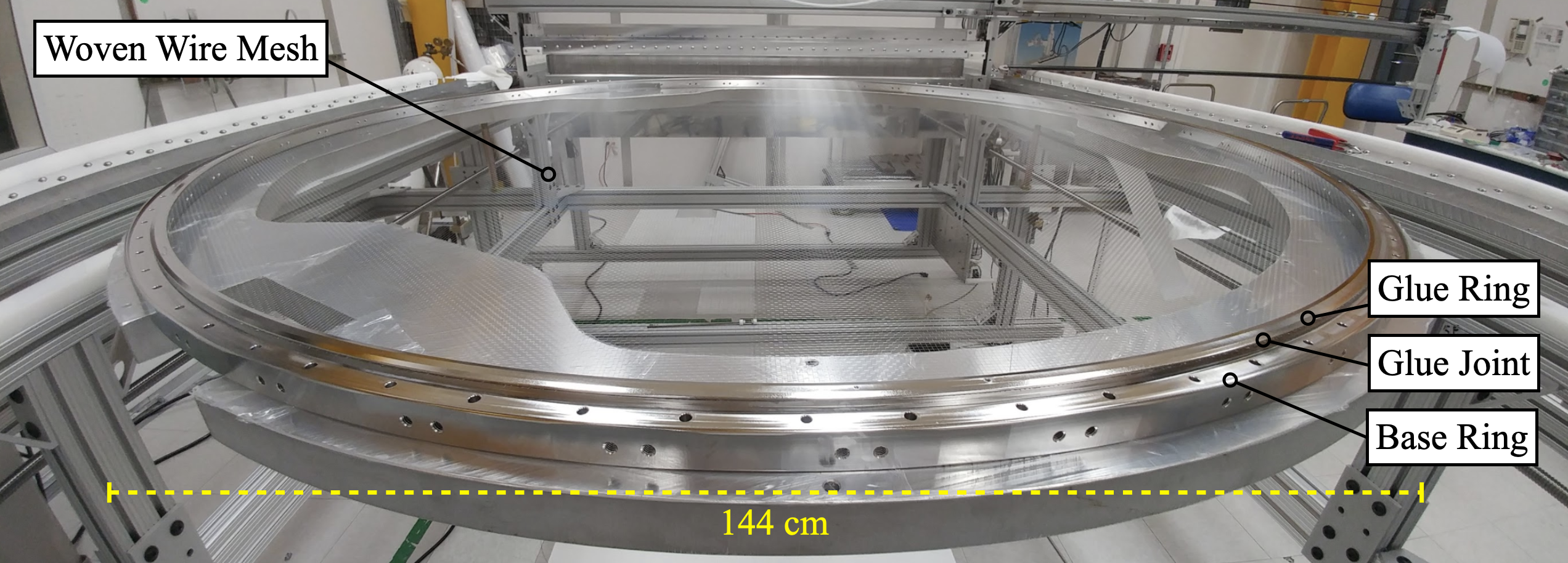}
\caption{A photograph of the anode grid, displaying the basic structure of an electrode grid. Note that the guard ring is not included in this photo.}
\label{AnodeOnLoom}
\end{figure*}

To produce useful electroluminescence signals, dual-phase xenon TPCs must maintain even higher fields at the liquid/gas interface and in the gas. Fields of at least 5~kV/cm just below the liquid surface are required for \(>\)85$\%$ efficiency of electron extraction from the liquid xenon into the gas~\cite{LivermoreExtractionEfficiency}. Achieving such a high extraction efficiency has an added benefit of reducing backgrounds due to delayed electron emission from the liquid surface~\cite{LUXEBackgrounds}. Once in the gas, the number of photons produced per unit distance traveled in xenon, $dN_{p}/dx$, can be parameterized by
\begin{equation}
\label{eq:dNdx}
    \frac{dN_{p}}{dx} = \alpha E - \beta P - \gamma,
\end{equation}
where \(E\) is the electric field, \(P\) is the pressure, and \(\alpha\), \(\beta\), and \(\gamma\) are empirical parameters given in Ref.~\cite{ChepelAraujo}. One can then see that the onset of electroluminescence in saturated xenon vapor at 1.8~bar(a) (the nominal detector pressure) occurs at a field of 2.7~kV/cm. To ensure a sufficiently large number of photons for reliable energy and position resolution, the fields in this region are typically held much higher, often around 6--8~kV/cm~\cite{ManninoThesis,X1TEnergyResolution}.

Maintaining drift and extraction field uniformity is critical for dual-phase TPCs to function properly. Drift field uniformity is key to accurately reconstructing the event location and simplifies the identification and rejection of events at the TPC walls. Since these wall events are overwhelmingly radiogenic in origin, this capability plays an important role in discriminating WIMP signals from background. Field uniformity in the electron extraction region is also beneficial, as nonuniformities introduce an \textit{xy} dependence to the electroluminescence yield and extraction efficiency. The imperfect correction of these spatial effects leads to degradation of the S2 (and therefore energy) resolution. Maximizing energy resolution is critical for increasing sensitivity to (0$\nu\beta\beta$) decay~\cite{DoubleBetaDecaySensitivity}.

While it is necessary to maintain strong and uniform drift, extraction, and electroluminescence fields, it is also ideal to minimize fields elsewhere in the detector as much as possible. This reduces sporadic, field-induced electron emission that can mimic or substantially distort S2s from low-energy scatters like CE\(\nu\)NS and low-mass WIMP recoils, where S2s are made from only a few electrons~\cite{BaileyThesis}. This is particularly important to consider for the grid wires, where the electric field strength can reach 50 kV/cm or higher, and where emission can vary significantly depending on microscopic surface defects that are hard to eliminate or control~\cite{BaileyThesis,ImperialWireStudies}. Minimization of stray fields also lowers the likelihood of electrical breakdown, in which electronic current between electrodes increases in an uncontrolled manner and risks damage to TPC components.

\subsection{The LZ High Voltage Grids}
The LZ TPC uses four high-voltage (HV) electrode grids to create the drift and extraction fields, as shown in Figure~\ref{TPCDiagram}. The gate, cathode, and anode positions and operating voltages achieve a nominal 0.3 kV/cm drift field between the cathode and gate and an extraction field of 5 kV/cm in the liquid (\(\sim\)10 kV/cm in 1.8~bar(a) xenon gas) between the gate and anode. The bottom grid is responsible for shielding the bottom PMTs from the strong (3.5 kV/cm), upward-pointing fields in the region below the cathode. A similar shield grid positioned below the top PMT array was considered but rejected. The field there is lower (1.0 kV/cm), reducing the need for such a grid. Using such a grid to lower the field at the top PMTs also would enhance the field below this grid, which is undesirable in the gas as it could enhance stray electroluminescence signals. Moreover, implementing such a grid in the gas phase would have added additional complexity to the mechanical and HV design of the above-anode region, making the cost of adding a fifth grid outweigh the grid's benefits.


Each grid is a mesh woven with thin stainless steel (SS304) wire. This mesh is attached to a set of rigid SS304 rings with a low-outgas epoxy designed for cryogenic applications (Figure \ref{AnodeOnLoom}). The radii of these rings were chosen to be as large as possible while still maintaining sufficient distance between their high voltage surfaces and the grounded ICV wall at a radius of 78.9 cm.

The grids will each deflect (`sag') due to electrostatic pressure from the fields in the various TPC regions. The anode and gate will deflect by roughly a millimeter toward each other, and the cathode and bottom will deflect toward each other less. A desire to limit these deflections and the resulting field nonuniformities motivated many of the design decisions for the grids. This was particularly important for the extraction region, where small, mm-scale grid deflections are non-negligible compared to the 1.3 cm undeflected separation of the gate and anode. In this region, such deflections change both the field strength in the gas and the distance over which extracted electrons travel to the anode, resulting in measurable changes to the size and shape of S2 pulses.

\subsection{HV Electrode Grid Design}

The most critical decisions made in the design of the HV electrodes were the selection of a woven wire mesh, the method of attaching the mesh to its rings, the design of the ring sets, the selection of a wire material, the selection of a mesh density, and the choice of a wire tension.


\subsubsection{Selection of Woven Wire Mesh}
Four potential electrode designs were considered for LZ: planes of parallel wires, crossed unwoven planes of wires, etched grids, and woven wire meshes~\cite{FNALHVWorkshop}.

Electrostatic considerations generally favored woven meshes for two reasons. First, studies of S2 resolution strongly favored the use of a spatially uniform structure for the anode plane, similar to the wire mesh anode used in LUX~\cite{BaileyThesis}. Wire meshes and grids chemically etched from thin foil~\cite{XENON100} were found to be superior to simple parallel wire planes with regard to maximizing electric field uniformity. Crossed wire planes were also disfavored here, as the planes would have deflected away from each other once the grid was taken to voltage, decreasing field uniformity. The second reason why meshes were electrostatically desirable was due to their smooth electrode surface. Etched grids were thought to be somewhat less advantageous due to the rectangular cross sectional shape of the metal elements. Sharp corners left by the etching process create higher fields on the electrode surface than what would be present on round wires, and are therefore less ideal for limiting electron emission. 

\begin{figure}[b!]
\includegraphics[width=\linewidth]{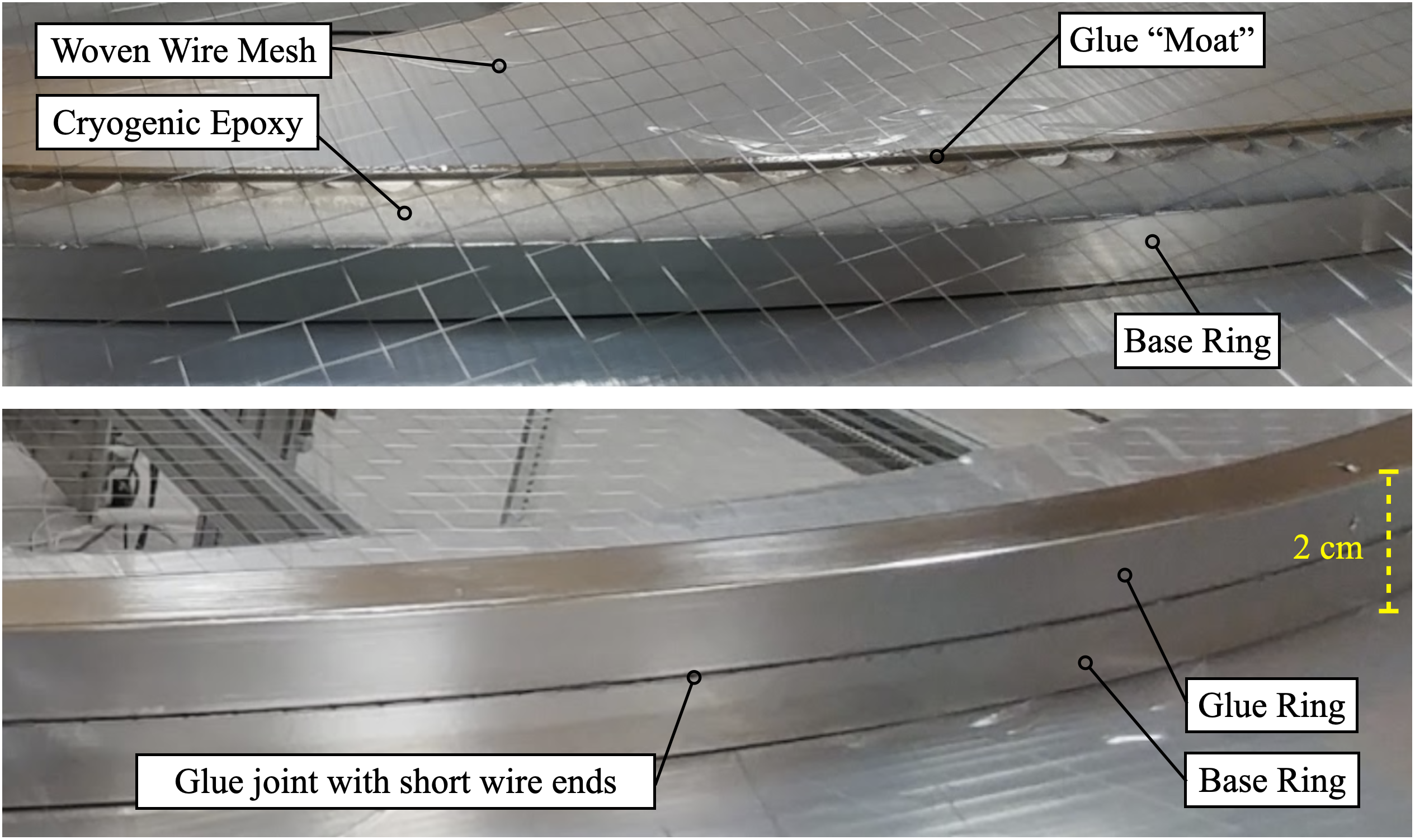}
\caption{Top: a photo of the epoxy connection between the cathode wires and base ring, taken just after glue deposition. Bottom: a photo of the cathode wires secured to the glue and base ring, taken after glue curing and grid cutdown.}
\label{WireAttachment}
\end{figure} 

Mechanically, woven meshes were also the superior design choice at LZ's 1.5 meter scale. Grid tension needed to be sufficiently strong and uniform to prevent excessive electrostatic deflection. Parallel wires strung only along one axis can, if tensioned sufficiently, introduce a substantial non-uniform load in the supporting ring. In such a scenario, rigid, high mass rings must be used to limit out-of-roundness. However, this increases the radiogenic background from the rings. It also increases the probability that the rings will obscure secondary scatters of gamma-rays and neutrons with insensitive ``dead'' mass, mass that displaces sensitive xenon where these interactions would otherwise occur. A design with unwoven crossed wire planes oriented at right angles to each other reduces the nonuniform load in the supporting ring, but also dangerously encourages a 3D saddle-like ring deformation. Moreover, if the common strategy of individually securing wires to the ring is used in either a crossed or parallel plane design, tensioning all wires uniformly is challenging. This task must be done iteratively, as changing an individual wire's tension can also change the tension in the others if the supporting rings are not sufficiently rigid~\cite{ManninoThesis}. Use of sophisticated wire-winding machines for tensioning, as was done for the ZEPLIN-III cathode~\cite{ZEPLIN-IIIDetector}, was possible but was deemed cumbersome at this scale for a circular ring. As a result, the parallel and crossed wire plane designs were disfavored. In contrast, woven meshes and etched grids provide a more practical means for achieving uniform and isotropic tension in the electrode. However, these designs initially proved difficult to scale to a 1.5~meter diameter electrode. Fine commercial meshes at this scale were unavailable. In addition, smaller commercially-available test meshes tensioned at the 1~meter scale tore at tensions far below those desired for use in LZ, and did not allow for convenient setting and assessment of tension throughout the grid. To overcome these obstacles, the decision was made to produce a mesh using a custom-built loom. In this way, each wire could be brought to a predefined tension of 2.45~N before being woven into the grid, ensuring that uniform tension was applied across the electrode at all times during and after the construction process. The woven wire mesh design was deemed suitable for all four of the LZ grids, and as a result all four grids were constructed on this loom.

\subsubsection{Mesh-Ring Attachment Method} The grid mesh-to-ring connections needed to be resistant to outgassing and cryogenically robust. Within these constraints, a connection with a smaller radial footprint was advantageous, because it would reduce the dead mass within the grid rings, and would help limit fields between the ring and the ICV wall. Lastly, the method of implementing the connection needed to be simple and resistant to human error. The last two constraints discouraged the use of individual wire-ring connections. Due to the woven mesh structure, such individual attachment points could substantially increase the size and design complexity of the rings. In the final design, a low-outgassing cryogenic epoxy from MasterBond (part no. EP29LPSP) was used to secure the wires in a single, continuous glue ``joint" between a lower (base) and upper (glue) ring (Figure \ref{WireAttachment}). Plastic (PMMA) spacer beads were added between the rings to prevent the glue ring from putting a direct load on the wires.\footnote{During development, wires broke under the glue ring prior to the introduction of the spacer beads. This was thought to be due to nonuniform load on the wire crosses introduced by tiny imperfections in the ring flatness.} These beads were 212--250~\microns\ (250--300~\microns) in diameter for the gate and bottom (anode and cathode) grids, with sizes based on the selected wire diameter for each grid (see Section \ref{subsubsec:MeshDensity}). The shear strength of the EP29LPSP epoxy was large enough that even for the narrowest glue surface, the wire tension required to break the glue joint was about 30 N, approximately ten times higher than the wire tensions considered for use with the grids.\footnote{This safety factor may have been lowered somewhat given the grids' non-standard cure schedule (described in Section~\ref{subsubsec:Gluing}), but it is not expected to vanish completely.} As a result, this epoxy connection simplified the ring design and safely permitted ring widths below 1 cm. While these advantages came at the cost of developing an automated glue dispensing system for reproducible application of epoxy (Section \ref{sec:GridProdEnvironment}), they were significant enough to motivate selection of the epoxy connection in the final electrode design.


\begin{figure}[]
\centering
\includegraphics[width=\linewidth]{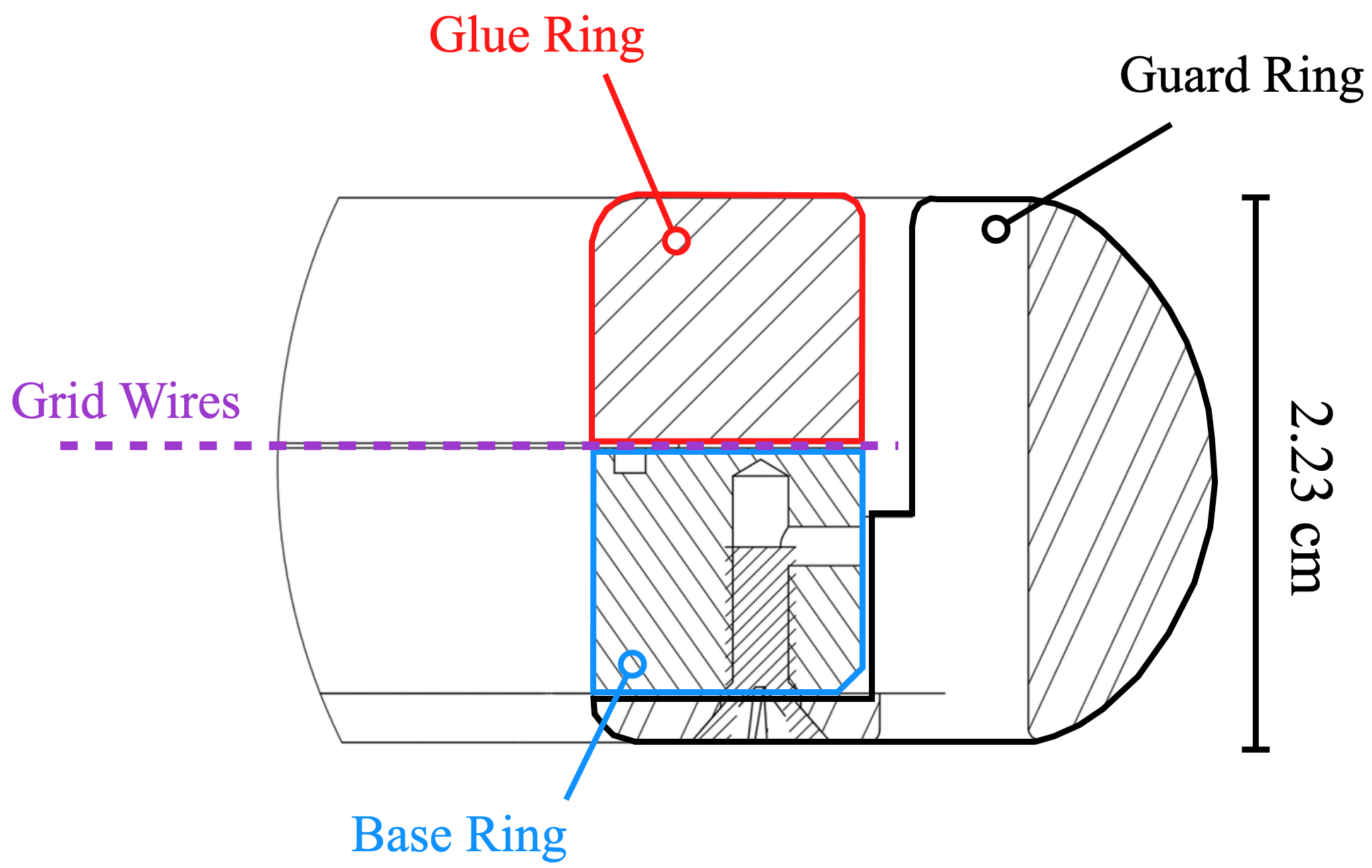}
\caption{A cross sectional view of the components of the cathode ring set, showing glue and base rings of equal thickness. The image shown does not represent all ring sets, but all ring sets have the basic three-ring form shown here. Cross sections of the gate and anode ring sets showing their large base rings and thin glue rings can be seen in Figures \ref{ERCutaway} and \ref{ComsolER}.}
\label{CathodeRingset}
\end{figure}

\subsubsection{Ring Set Design} 

The grid ring sets were designed in part to accommodate the selected mesh-ring attachment procedure. Each was built out of three rings: a ``base" ring, a ``glue" ring, and a ``guard" ring, as shown in Figure~\ref{CathodeRingset}. A small trough (the ``moat") was included on the inside of the base ring to capture any excess glue attempting to flow inward toward the grid. The guard rings were designed to create field-free regions for any small wire ends protruding outward from between the glue and base rings. Without this design, fields on these ends could extend well into the range of MV/cm, leading to electrical breakdown or electron field emission. 

\begin{table*}[t!]
\centering
\caption{Properties of materials considered for grid wire. Coefficients of thermal expansion (CTEs) shown in this table are taken from Refs.~\cite{asm} and~\cite{engineeringtoolbox}. UTS and YTS values are taken from Refs.~\cite{cfw}, and are not available for all wire material choices. VUV optical reflectivity values are taken from Refs.~\cite{copperReflectivity} and \cite{goldReflectivity}, \cite{alReflectivity}, and~\cite{ssReflectivity}. Those including ``\(<\)'' are upper bounds, reflecting the fact that oxidized surfaces or those with an imperfect finish will have a lower reflectivity. The tungsten (W) YTS is left blank because the value quoted by California Fine Wire for various work-hardening levels of tungsten is not well-measured.}
\begin{tabular}{|p{2.3cm}||p{1.0cm}|p{1.7cm}|p{2.0cm}|p{1.7cm}|p{2.5cm}|}
\hline
Wire Material & UTS (psi) & 0.2\(\%\) YTS (psi) & CTE (\microns/(m$\cdot $K)) & Work Function (eV) & VUV Optical Reflectivity (\(\%\)) \\
\hline
\hline
SS304 & 360,000 & 280,000 & 17.3 & 4.3 & \(<\)57 \\
SS302 & 300,000 & 280,000 & 17.2 & 4.3 & \(<\)57 \\
SS316L & 245,000 & 230,000 & 16.0 & 4.3 & \(<\)57 \\
BeCu Alloy 25 & 150,000 & 125,000 & 17.8 & 4.65 & 20 \\
W (Au) & 600,000 & -- & 4.5 & 5.1 & 23 \\
Al 5056 & 63,000 & 59,000 & 22.5 & 4.28 & \(<\)90 \\

\hline
\end{tabular}
\label{table:wireProperties}
\end{table*}

Each ring set had unique design features according to the grid's role in the detector. The bottom and cathode grids were generally kept as light as possible to reduce their radioactive footprint and dead mass. The cathode ring's OD was kept as small as possible to maximize the distance from the ring to the grounded ICV wall and thus minimize the field in between. Because the cathode and bottom rings were fairly light, the plane of the wire mesh was designed to be near the vertical center of the ring set to minimize the wire tension's tendency to twist the rings around the axis perpendicular to the ring cross section. This led to glue and base rings of nearly equal height. In contrast, large, high-rigidity base rings were needed to provide mechanical support for the extraction region. This rigidity easily resisted torsional forces of the wires, allowing the glue and guard rings to be made very thin. The glue rings for the anode and gate therefore served more as a well-defined equipotential and buffer between xenon and epoxy than a structural addition to the grid. This design with minimally thin glue rings was also necessary to minimize the field between the anode and gate ring surfaces. Keeping this field low was critical to suppressing gate ring field emission and any spurious ``S2-like" signals that it could create. SS304 was chosen as the material for all rings, as it was relatively simple to machine (compared to titanium, for example) but didn't compromise on strength.


\subsubsection{Wire Material}The wire material was selected by weighing the optical, thermal, electrical, tensile, and radiogenic properties of the candidate materials. Table~\ref{table:wireProperties} compares many of these properties for several different materials. 

Optical raytracing simulations demonstrated that within the highly reflective TPC, the grid wire parameters are important in determining the overall light collection efficiency (LCE)~\cite{conceptualDesignReportLZ}. This fact favors smaller wire diameter or larger pitch. However, the wires must also be tensioned to limit electrostatic deflection, which favors a thicker wire or smaller pitch. To limit deflection without decreasing the LCE, one can select a stronger wire material. Stainless steel 304's (SS304) high yield (YTS) and ultimate (UTS) tensile strengths best accomplished this balance. Moreover, even though the reflectivity of SS304 for the VUV luminescence of xenon near 175~nm varies depending on the surface finish, it is expected to be higher than that of other common wire materials or coatings, such as copper or gold~\cite{copperReflectivity}~\cite{goldReflectivity}. While aluminum has a higher reflectivity at these wavelengths, its tensile strengths are significantly lower than those of SS304, making it challenging to limit electrostatic deflection without degrading LCE.

Thermal, radiogenic, and electrical considerations also motivated the final selection of SS304 wires. SS304 wires matched the coefficient of thermal expansion (CTE) of the grid rings, which were also SS304. As a result, no change in wire tension would be expected after cooling the TPC to liquid xenon temperature. Simulation studies also showed a slight preference for SS304 over BeCu wires due to the much lower rate of (\(\alpha,n\)) interactions within the wires. Though its electrical performance and work function were not as promising as those of BeCu and gold-coated tungsten wires, there was a comparatively broader level of experience with its use in low-background xenon TPCs. This familiarity in part made this material selection tolerable from the electrostatic performance perspective. SS316L was also known to have high corrosion resistance and showed additional robustness to electron emission over SS304~\cite{ImperialWireStudies}, but this fact had not been fully recognized by the time the final wire had been procured. In the final design, a ``hardened" form of SS304 with the highest possible UTS was purchased from California Fine Wire in Grover Beach, CA, for use in the LZ grids.





\begin{table*}[t!]
\centering
\caption{Average values of the (undeflected) grid wire surface fields, per Equation \ref{eq:GaussLaw}. Due to the nonuniform charge distribution around a cylindrical wire in a parallel plane, the maximum surface field is actually between 5\(\%\) and 8\(\%\) higher than the average value from Equation 3 for LZ's grids. The woven-mesh design adds an additional variation to the wire surface charge density. For the gate, the maximum wire surface field is 30\(\%\) higher than the quoted average field. For the anode, the maximum is 40\(\%\) higher. In both cases, the maximum field occurs at the farthest point from the mesh wire crossings, on the side of the wire facing the liquid surface.}
\begin{tabular}{|P{1.5cm}||P{1.0cm}|P{2.6cm}|P{2.6cm}|P{1.7cm}|P{3.5cm}|}
\hline
Grid & Pitch (mm) & Wire Diameter (\microns) & Average Surface Field (kV/cm) & Cathodic? & Optical Transparency (Normal Incidence) \\
\hline
\hline
Anode & 2.5 & 100 & 43.9 & No & 92\(\%\) \\
Gate & 5 & 75 & 49.9 & Yes & 97\(\%\) \\
Cathode & 5 & 100 & 30.6 & Yes & 96\(\%\) \\
Bottom & 5 & 75 & 34.0 & No & 97\(\%\) \\
\hline
\end{tabular}
\label{table:MeshDensity}
\end{table*}

\subsubsection{Mesh Density}
\label{subsubsec:MeshDensity}The selection of each grid's mesh density involved a balancing of optical performance against electrostatic performance. For a given bulk field between grids, a larger pitch or smaller wire diameter requires more field lines to be concentrated on an individual wire's surface, raising the surface field and increasing the risk of electron emission. However, optical transparency also increases with a larger pitch or smaller diameter, which increases light collection.

More formally, a square grid with pitch $s$, wire diameter $d$, and a wire reflectivity of 0$\%$ has an opacity to normally incident light given by
\begin{equation}
    \frac{2sd-d^{2}}{s^{2}} \simeq \frac{2d}{s}\ ,\  d \ll s.
\end{equation}
For wire diameters that are small relative to the pitch, the opacity scales as \(d/s\).

In contrast, the wire surface electric field scales as $s/d$. A simple estimate of the average wire field can be found by approximating a woven grid with pitch $s$ as a single plane of parallel wires with pitch $s/2$. Applying Gauss's Law to a wire ``cell'' gives
\begin{equation}
\label{eq:GaussLaw}
    \langle E_{s}\rangle \ =\ \frac{s}{2\pi d}[E_{z,+}-E_{z,-}],
\end{equation}
where $E_{z,+}$ and $E_{z,-}$ are the \textit{z}-components of the bulk fields above and below the grid, respectively.\footnote{The logic behind this approximation is that a crossed wire grid with pitch $s$ and a parallel wire plane with pitch $s/2$ have the same area of metal on which electric field lines can terminate. This implies that for a given bulk field strength, the density of field lines at the wire surface will be similar. For more details on deviations from this model, see the caption to Table \ref{table:MeshDensity}.}

A mesh density was selected for LZ by maximizing the optical transparency of the grid while keeping low enough surface fields to limit electron emission. Motivating limits on surface fields from a theoretical standpoint was challenging. Historically, electron emission was thought to be primarily due to Fowler-Nordheim (F-N) tunneling induced by electric field enhancements around asperities on the wires~\cite{BaileyThesis,fieldEmissionTheory,Williams_1974}. This model would lead in practice to a sharp, monotonic rise in electron emission over a narrow range of wire surface fields. However, for SS304 in liquid xenon, the sharp rise is theoretically predicted to begin at cathodic surface fields on the order of 10 MV/cm. These surface fields are much higher than those at which electron emission rates have spiked in past detectors, even after making generous assumptions about the geometry of any possible field-enhancing asperities on those wires. Moreover, recent studies have cast doubt onto traditional F-N tunneling as the primary mechanism for the field-induced electron emission seen in liquid xenon detectors, instead suggesting that emission is primarily dependent on the state of corrosion products on the surface of the wire~\cite{ImperialWireStudies}. In light of this uncertainty, an upper limit on cathodic surface fields in LZ was set by surveying the maximum surface fields achieved in stable operation of recent liquid xenon detectors using stainless steel wire~\cite{conceptualDesignReportLZ}. Those detectors were able to run at surface fields between 40 and 60~kV/cm. As a result, the maximum allowable cathodic surface field imposed for LZ was set to 50~kV/cm.\footnote{In LZ, this limit is achieved for a no-deflection scenario, but is exceeded when the gate and anode deflect toward each other. However, the possibility of additional wire treatments for reducing electron emission (Section \ref{subsec:Passivation}) justified the decision to accept this excess.} This criterion helped constrain the mesh density for the cathode and gate grids to the values shown in Table \ref{table:MeshDensity}.

While optical transparency was an important factor in selecting the anode and bottom mesh densities, these mesh densities were also chosen to achieve field uniformity in the extraction and reverse field regions. In particular, S2 resolution is sensitive to both the extraction region field uniformity and the anode wire surface fields (see section \ref{subsubsec:S2Design} for further discussion). To boost field uniformity, the anode pitch was selected to be finer than the other grids, and its wire diameter was selected to be relatively wide, though the selections were constrained by a desire to prevent excessive degradation of optical transparency. These selections also kept the anode and bottom average surface fields below 50~kV/cm. Even though these grids are not cathodic and are therefore not prone to electron emission, keeping surface fields sufficiently low reduces the risk of anode-initiated breakdown, improving the general high voltage performance of the TPC~\cite{plasmaChemistry}.



\subsubsection{Wire Tension}
\label{subsubsec:WireTension}The mesh wire tension was motivated primarily by the desire to limit the combined electrostatic deflection of the gate and anode to 2.0~mm. While it is not practical to limit deflection to the extent that S2 variation across the TPC is negligible, a 2.0~mm combined limit ensures that the ultimate variation in S2 is modest, and can be corrected in analysis. This limit corresponds to a maximum of 15\(\%\) decrease in the width of the extraction region, 11\(\%\) increase in the gate wire surface fields, and a 5\(\%\) increase in the number of photons produced per extracted electron at the center of the detector. Combining this effect with radial dependence of the LCE results in a nonuniform single electron (SE) pulse area that varies by 27\(\%\) from the edge to the center of the TPC. Analytical and numerical modeling of wire deflections motivated 2.45~N as the design tension for all grid wires. This model treated the grid as a single plane of parallel wires as in Section~\ref{subsubsec:MeshDensity} and used the net force on a wire from the fields above and below the grid to estimate the profile of that wire in equilibrium. This model was also expanded to include the feedback effect of deflection causing additional field enhancement (and further deflection) for the gate-anode system. The 2.45~N tension is expected to limit the combined gate-anode deflection to 1.4~mm while simultaneously keeping the wires a safety factor of more than 3 from their expected yield tensions.


\subsection{Electron Extraction Region Design}

\begin{figure}[t!]
\centering
\includegraphics[width=3in]{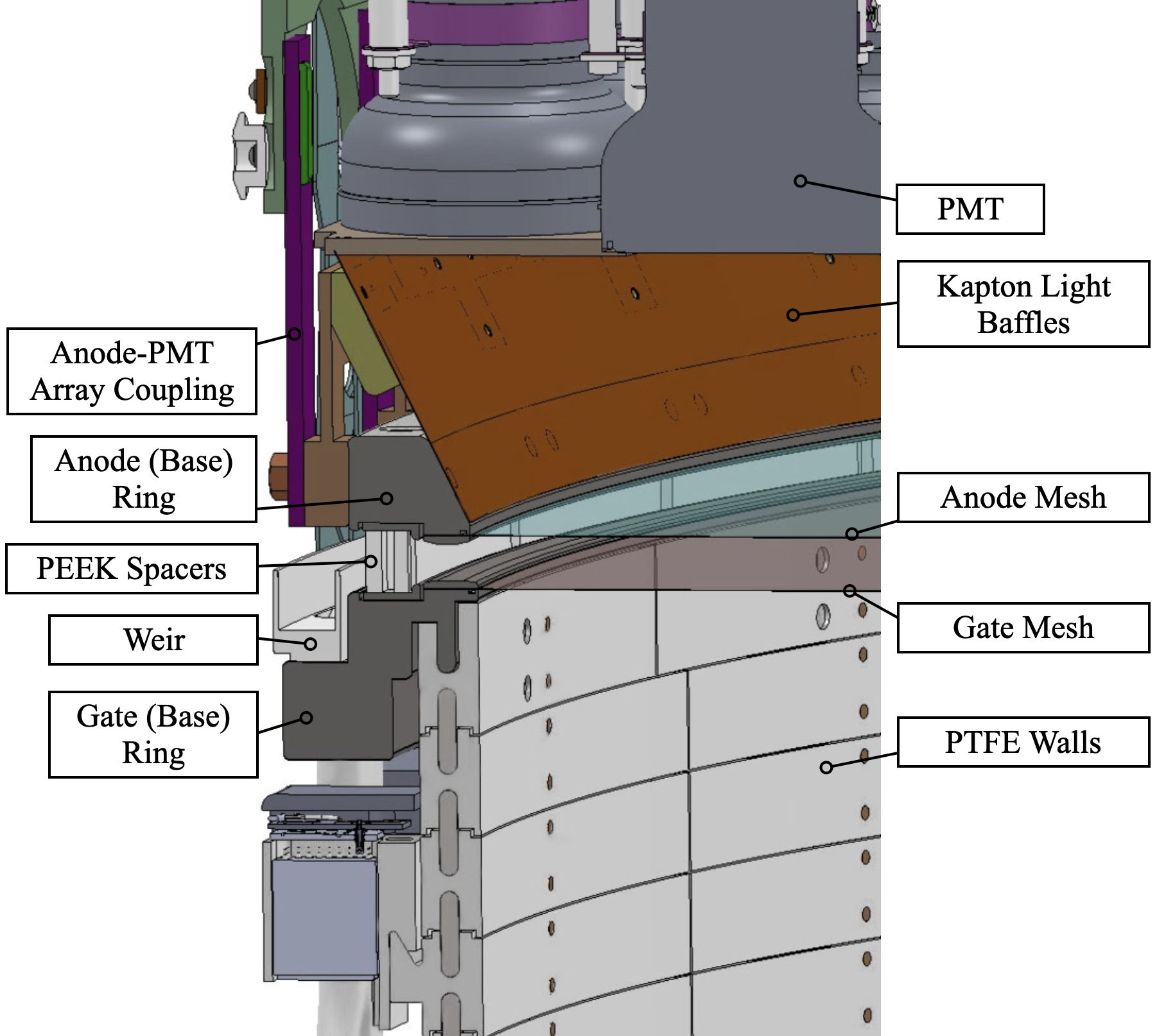}
\caption{A cross-section of the extraction region showing its place in the TPC.}
\label{ERCutaway}
\end{figure}

Tightly coupled to the design of the grids is the design of the electron extraction region (Figure~\ref{ERCutaway}). In our nomenclature, the ``extraction region" (ER) is defined as the gate grid, anode grid, and the structures in between. In this section, we discuss both this region and that between the anode and the top PMT array. Collectively, these regions are responsible for creating an optimal S2 signal, providing mechanical strength to the TPC, optically isolating the TPC from the xenon skin, and setting the liquid level.

\subsubsection{Mechanical Design} The extraction region was designed to be mechanically very rigid, so that the gate-anode grid distance would not be affected by potential non-uniform loads during handling, after ER integration into the TPC, or during detector cooling. This was accomplished through two aspects of the design. First, the gate and anode grid rings were designed with large cross sectional areas, ensuring that they would not distort much during handling or cooling. The second design choice involved the coupling between the rings. The gate and anode are separated by a set of 48 polyether ether ketone (PEEK) standoffs 13.6 mm tall (Figure \ref{ERCutaway}). They are aligned relative to each other by a set of 48 PEEK dowel pins (not visible in Figure \ref{ERCutaway}) that pass through the standoffs. The two grids are held together by 48 PEEK bolts, also passing through the standoffs. This coupling tightly constrains the two rings to each other, limiting changes to the extraction region as a whole when cooling.

Unlike most other pieces in and around the ER, the PEEK pieces connecting the anode to the PMT array were designed to retain some level of flexibility. This was done to provide a robust coupling that would not fail due to differential thermal contraction between the titanium structure housing the PMT array and the SS304 anode.

\begin{figure}[b!]
\centering
\includegraphics[width=\linewidth]{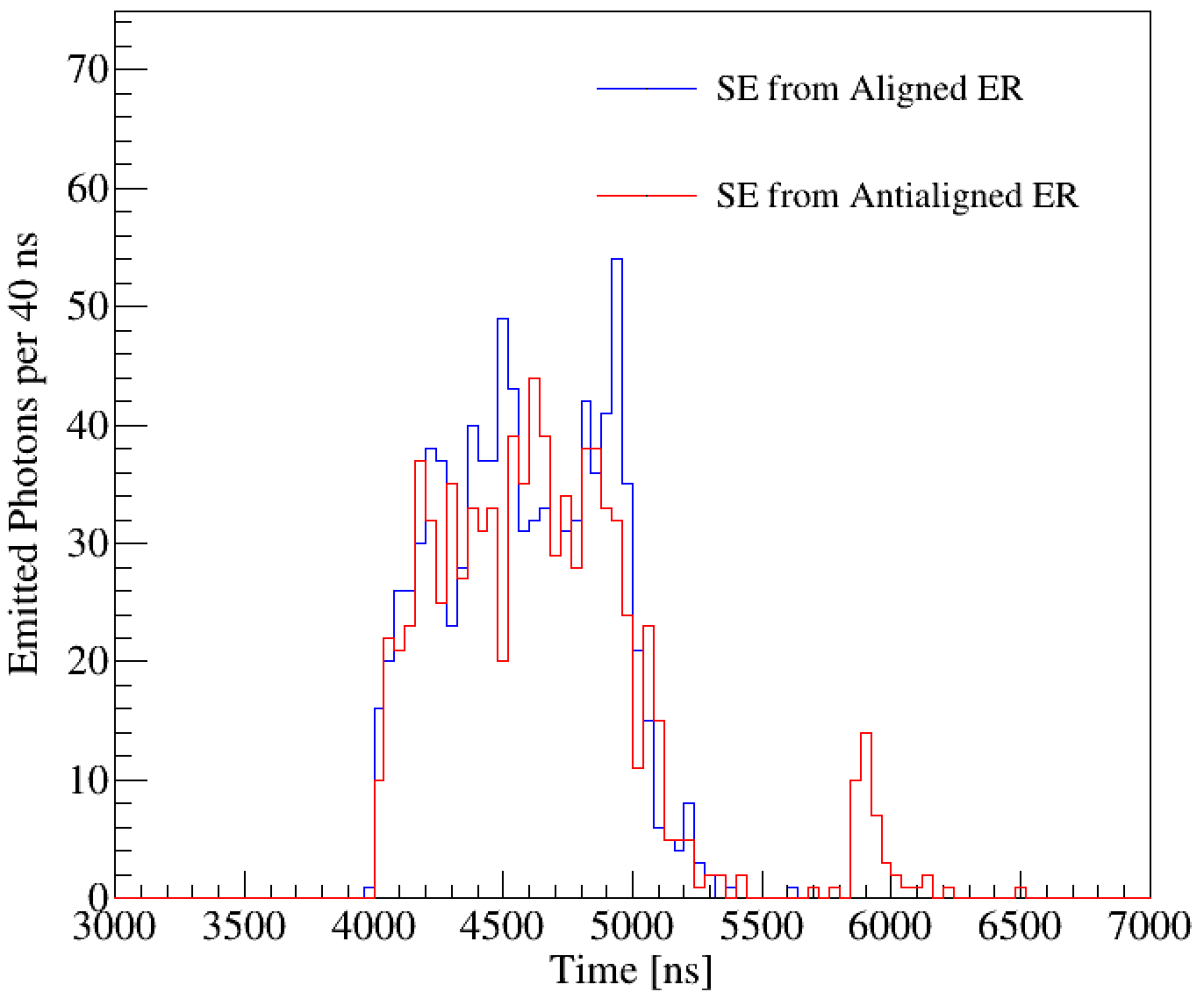}
\caption{Emission times of electroluminescence photons from two electron tracks simulated with the Garfield++ package~\cite{garfieldpp}. In this simulation, the AvalancheMC method outfitted with LXe drift parameters was used to transport each electron from 5~mm below the gate (T=0~ns) to the liquid surface, and the AvalancheMicroscopic method was used to drift the electrons to the anode in gas. These illustrate the difference between tracks that curl over the anode wires (red) and those that reach the wires from below (blue).}
\label{SETracks}
\end{figure}

\subsubsection{S2 Design}
\label{subsubsec:S2Design} Three requirements on the S2 response drove much of the design of the extraction region. First, the electroluminescence response needed to be robust enough to permit the detection of single electrons (SEs) with a high signal-to-noise ratio. Second, efficient (\(>85\%\)) electron extraction from the liquid was required, leading to a minimum field of \(\sim\)10~kV/cm just above the liquid surface. Third, a large SE pulse area at the edge of the detector was required to maintain good \textit{xy} reconstruction and rejection of small-S2 backgrounds near the walls, among which are decays of Rn daughters plated out onto those walls~\cite{tdr} and electron emission processes whose intensity correlates with the local event rate \cite{LUXEBackgrounds}. This requirement motivated a lower limit of about 40 photons detected (phd) for the area of peripheral SEs. An upper limit to the SE area was imposed by requiring a linear response for the PMT in the top array with the most electroluminescence light from a \(^{83m}\)Kr internal calibration source event. A final anode-gate gap of 1.3~cm, liquid-gate gap of 0.5~cm, and gate/anode voltages of -5.75/+5.75~kV/cm largely satisfied these requirements.

As described in Section~\ref{subsubsec:WireTension}, the SE size is expected to vary from center to edge of the detector. Assuming zero deflection at the nominal anode-gate \(\Delta V\), the electroluminescence yield is approximately 870-920 photons per electron, and is 5\(\%\) higher at the center, where the grids deflect. Accounting also for the varying LCE, the expected SE area is 98~phd at the TPC center and 76~phd at the TPC edge. Even considering the lower SE area around the walls, these areas are large enough to enable robust position reconstruction in low-energy WIMP search analyses, where events with as few as 5 electrons may be considered.\footnote{These large areas may give slight PMT saturation for \(^{83m}\)Kr decays, especially in the center of the TPC where the LCE is larger. However this saturation is currently anticipated to remain at the few percent level.}

\newcommand{\Garfieldpp}{Garfield\nolinebreak\hspace{-.05em}\raisebox{.4ex}{\tiny\bf +}\nolinebreak\hspace{-.10em}\raisebox{.4ex}{\tiny\bf +}}

\begin{figure}[b!]
\centering
\includegraphics[width=\linewidth]{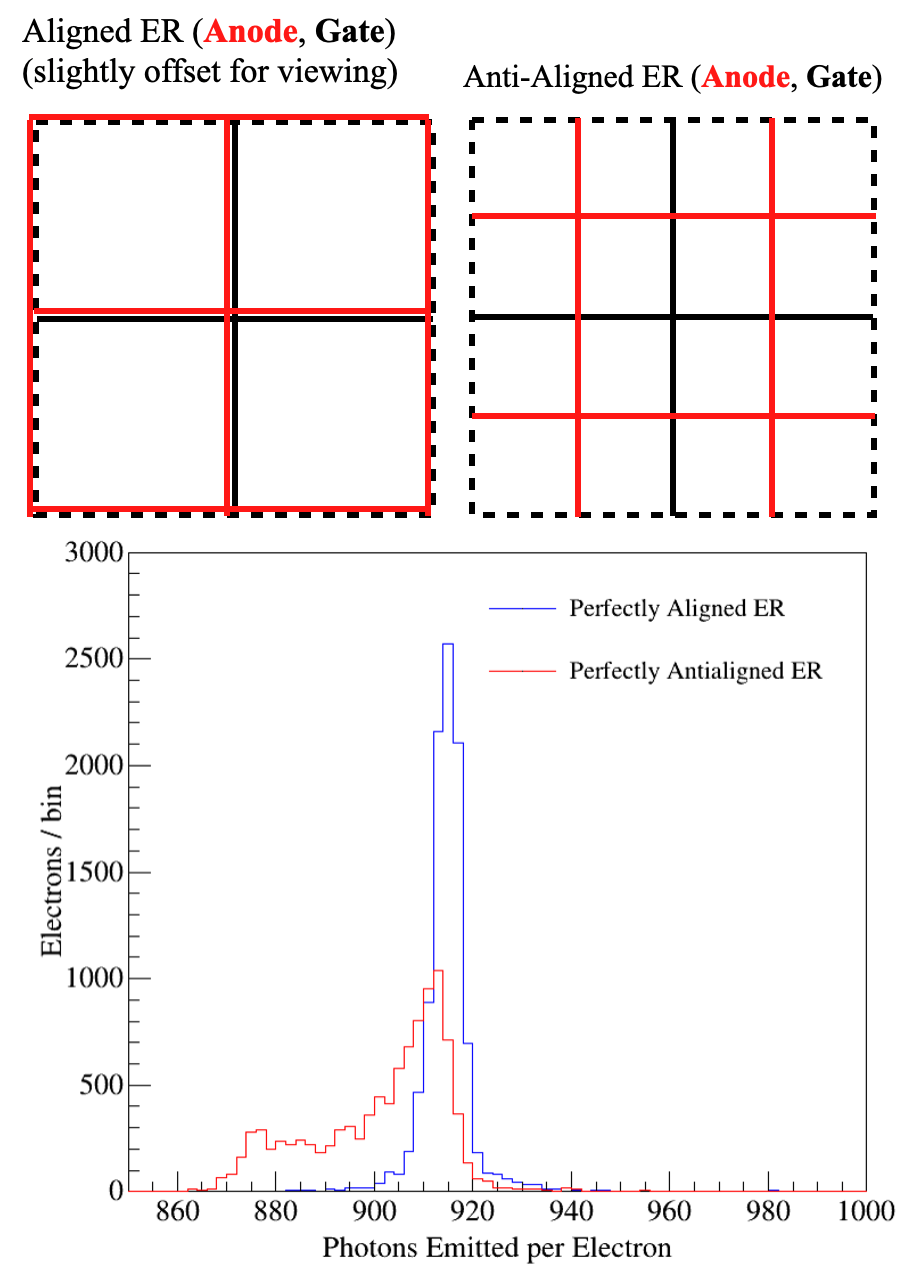}
\caption{Difference in photon yield between the aligned (top left) and anti-aligned (top right) anode-gate system. The plots are generated by simulating 10,000 electrons uniformly in space 5~mm below the gate wires, and using the Garfield++ software to generate electroluminescence from these electrons above the liquid surface~\cite{garfieldpp}.}
\label{Alignment}
\end{figure}

Maximizing S2 resolution for a 0\(\nu\beta\beta\) decay search drove other aspects of the ER design. An overall energy resolution of \(<\)2\(\%\) at 2.458~MeV was established as a requirement for LZ. The goal energy resolution was set at 1\(\%\), which has been achieved by a recently retired dual-phase xenon TPC~\cite{X1TEnergyResolution}. This resolution can be degraded by the grids in two ways. In the first, smearing occurs due to fluctuations in the gain process present when Townsend avalanches occur in the vicinity of the wires. These fluctuations are larger for a smaller wire diameter and a larger grid pitch. To limit degradation of S2 resolution from this effect, the anode was made a woven mesh with a tight, 2.5~mm pitch and a relatively large, 100~\microns\ wire diameter~\cite{BaileyThesis}. This selection maximizes field uniformity and results in essentially no multiplication near the anode wire surfaces. The second mechanism of S2 smearing is driven by electrons following drift paths to variable locations on the anode wires. Due to their longer tracks, drift paths for electrons that overshoot the anode before returning to the top of the mesh generate fewer photons per Equation \ref{eq:dNdx} than those ending on the bottom. This drift path variability can also affect the SE pulse shape in a way that may make it difficult to properly reconstruct the full pulse area. While all SEs have a spike at late times due to the electron entering a high-field region near the wire surface, drift paths terminating near the top of the anode wire introduce a large gap between the bulk of the SE and this late spike (Figure \ref{SETracks}).

The degree of alignment between the anode and gate grid cells (Figure \ref{Alignment}) determines the likelihood that electrons will curl around the anode. To assess the frequency with which this above-anode curling occurs, a custom addition to the \Garfieldpp\ electron drift package was made to propagate electrons using a set of field-dependent drift velocities and diffusion constants for liquid xenon based on work in Ref~\cite{Boyle}.\footnote{Code for adapting this technique to other phases or noble liquids can be found as part of the NEST package~\cite{NEST}.} Together with the COMSOL Multiphysics\(\textregistered\) electrostatics package, this permitted computational studies of drift and diffusion within complex ER geometries including both gas and liquid. Results from these studies show that a perfect alignment between the two grids discourages such paths and gives a tighter SE spread than perfect anti-alignment. For larger S2s, the spread in S2 size varies as a function of depth for a given ER alignment, being much larger for events occurring close to the gate than for those close to the cathode. For 2.458 MeV $^{136}$Xe $0\nu\beta\beta$ decays, the perfectly aligned (anti-aligned) extraction region gives a depth-averaged spread in generated S2 photons of 0.02$\%$ (0.1$\%$). To limit the S2 spread, the extraction region grids were designed to be aligned, rather than anti-aligned. Section 4 details how successfully alignment was accomplished during production.




\begin{figure}[b!]
\centering
\includegraphics[width=\linewidth]{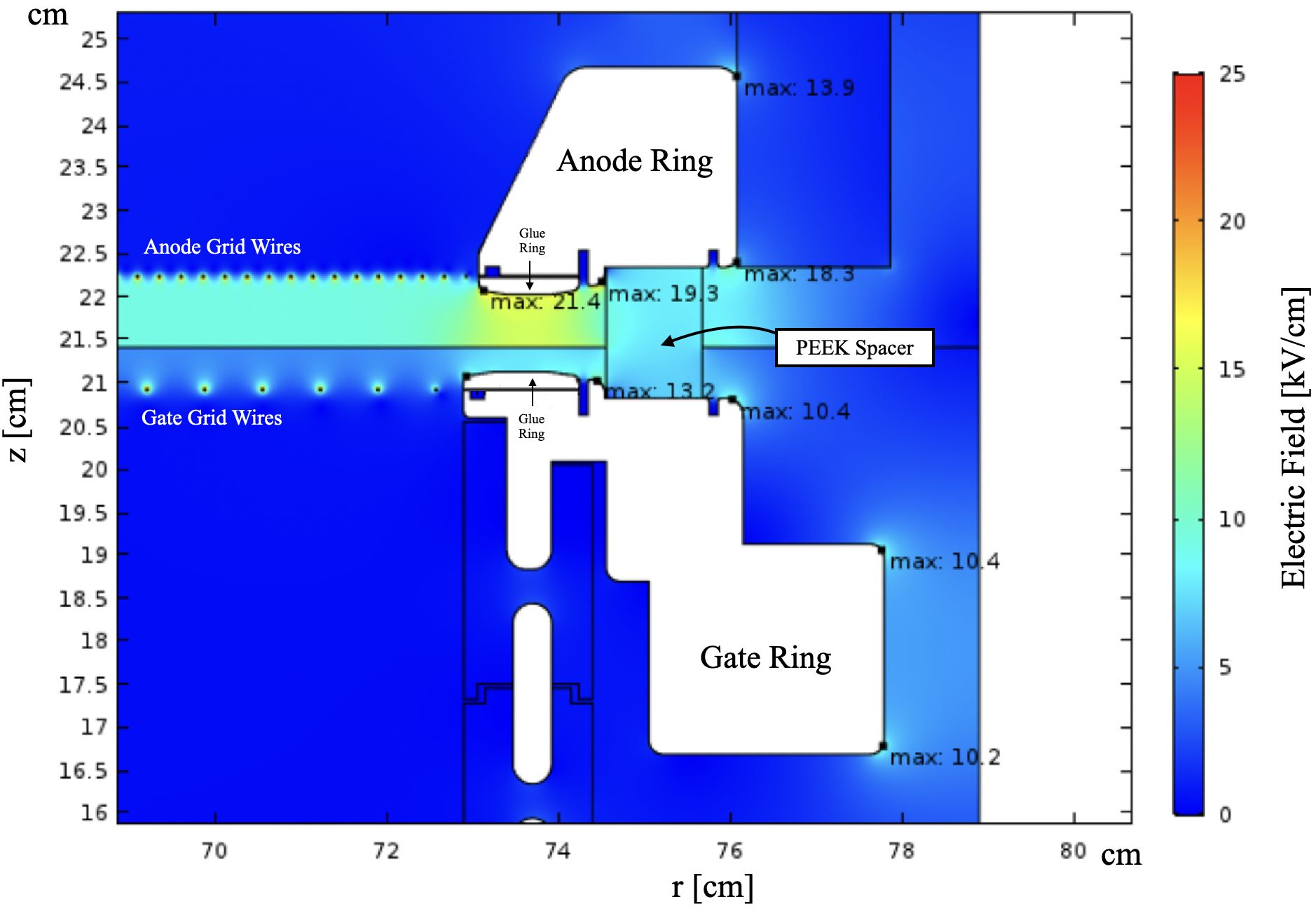}
\caption{A COMSOL Multiphysics\(\textregistered\ \)~axisymmetric 2D simulation of the electric fields in the extraction region and surrounding area. Numerical field values for a few key high field regions are computed and displayed.}
\label{ComsolER}
\end{figure}

\subsubsection{Peripheral Electric Field Design} The extraction region was also designed to minimize the likelihood of sporadic electron emission and breakdown from the rings to each other or to the grounded ICV wall. An anticipated weak point was the interface between the PEEK spacers and a given grid ring (Figures \ref{ComsolER} and \ref{spacerLabyrinth}), where a triple junction point between metal, PEEK, and xenon could substantially enhance electric fields by a factor of 2--3. To reduce the likelihood of electrical breakdown from this effect, the PEEK spacers were not placed directly between the gate and anode glue rings, where the field was highest. They were instead moved radially outward into regions of lower field between the rings. The glue rings were designed with a special rounded profile to maximize radii of curvature and minimize the fields on the ring corners.

\begin{figure}[]
\centering
\includegraphics[width=7cm]{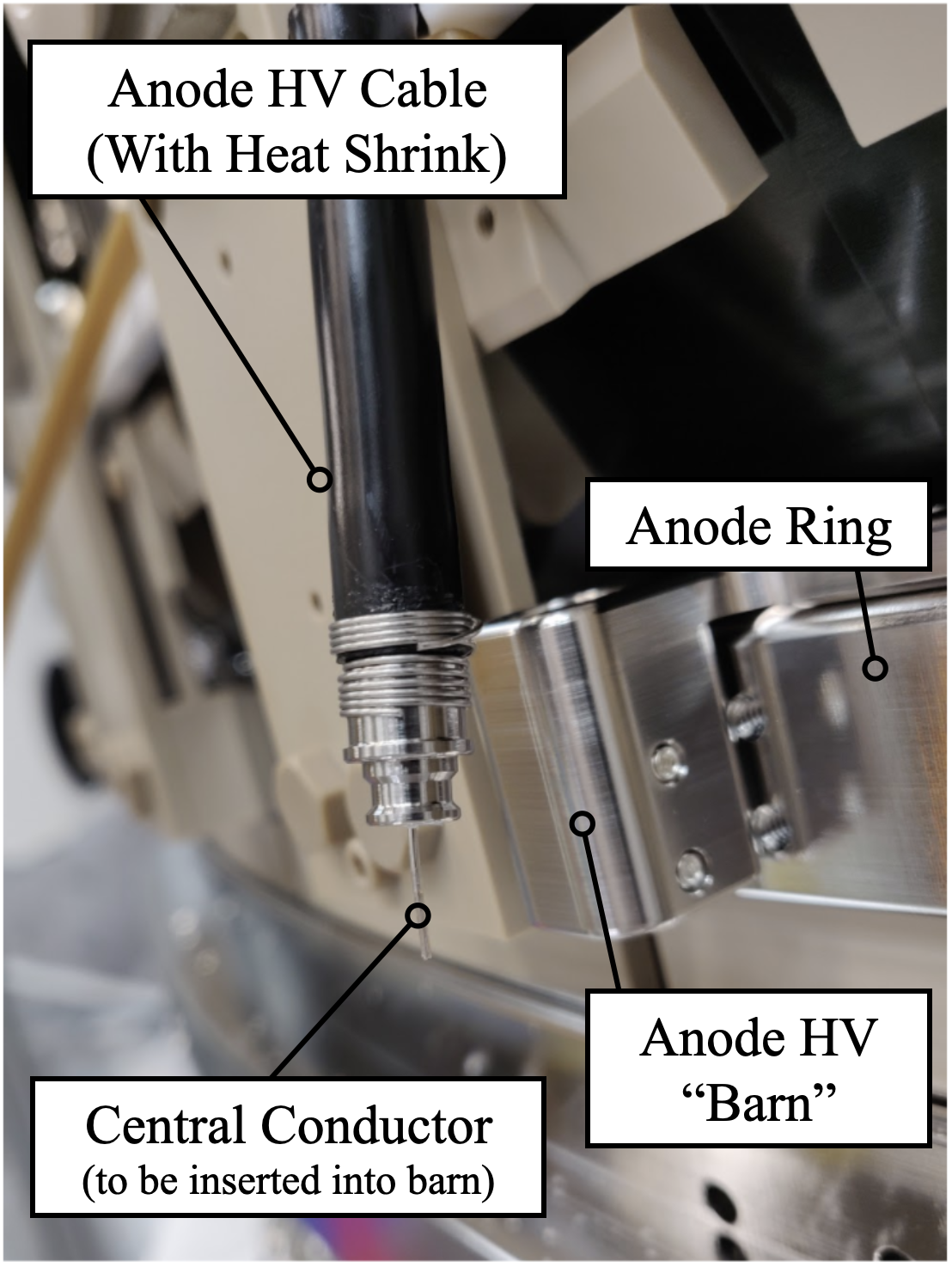}
\caption{A photograph of the HV cable connection for LZ's anode grid.}
\label{fig:anode_hv}
\end{figure}

A design and testing campaign was performed to optimize the gate and anode HV cables and cable-ring connection. This was done to mitigate the risk of breakdown along the cables or between the cables and other components of the detector. The final cables each had an Ag-plated Cu-clad steel central conductor separated from an Ag-plated Cu ground braid by PTFE. Fluorinated ethylene propylene (FEP) composed the jacket. The central conductor of the anode cable was inserted into a stainless steel ``barn" attached to the anode ring (Figure \ref{fig:anode_hv}). Resistive heat shrink added to the end of the cable provided a well-defined field grading down to the ground braid and reduced breakdown hazards from sharp wire ends on the braid. The same design was used for the gate connection, though its barn had a slightly different geometry because of the tight space in the skin region. Due to these space constraints, a PEEK cable router was used to constrain the path of the gate HV cable to minimize the likelihood of breakdown.

\subsubsection{Optical Design} 
\begin{figure}[]
\centering
\includegraphics[width=\linewidth]{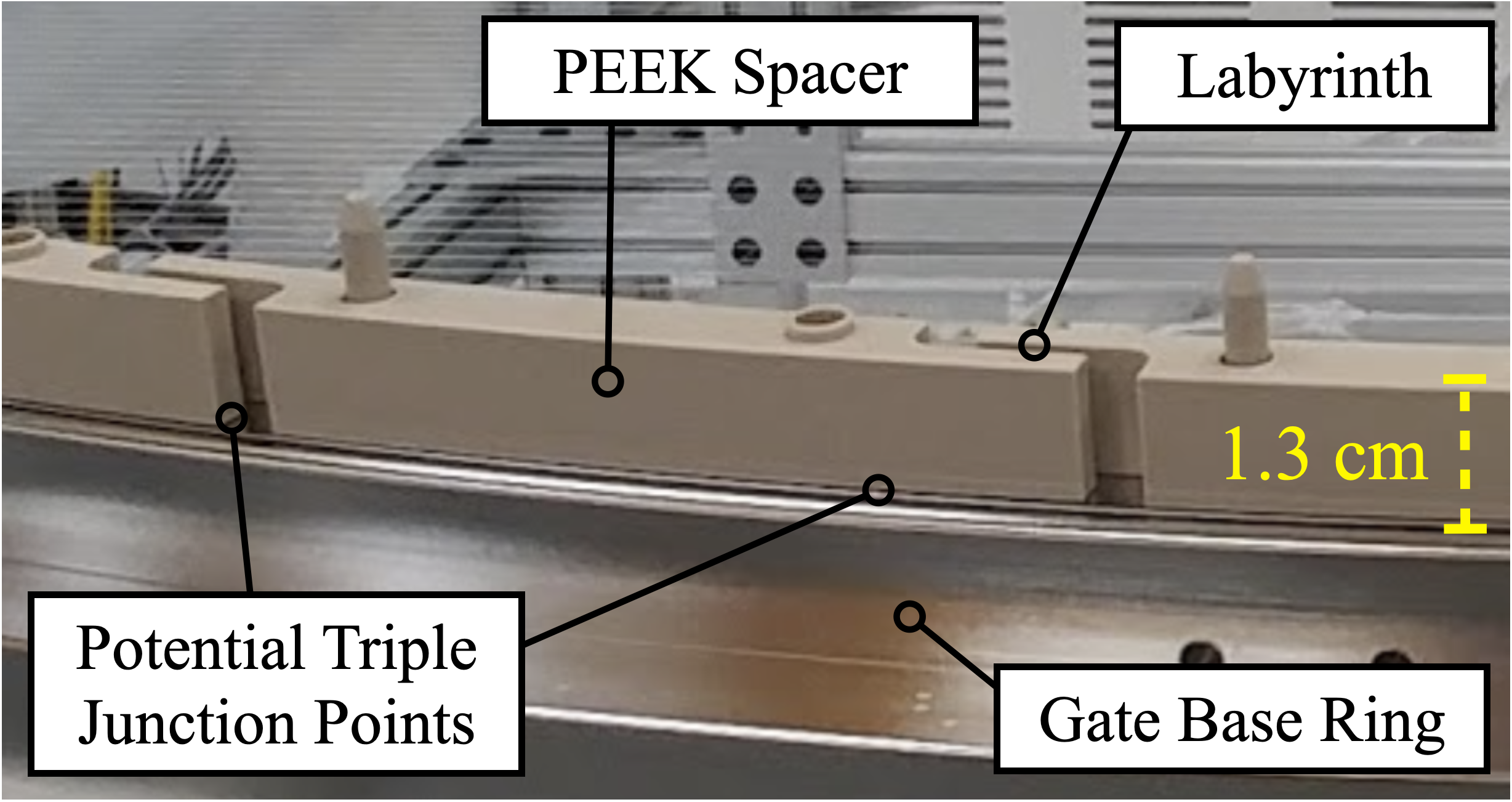}
\caption{The labyrinth structure in the PEEK spacers separating the gate and anode. This photo also notes the points where PEEK spacers met the grid ring to create a triple junction point, enhancing fields. While only the interface of the spacer and the gate ring is pointed out for visual clarity, the interface between the spacer and the anode ring also caused field enhancement.}
\label{spacerLabyrinth}
\end{figure}

The extraction region was designed to optically decouple the xenon skin region from the TPC. If stray S1 light in the skin region is allowed to reach TPC PMTs, it may pile up with an uncorrelated S2 and mimic a low-energy WIMP scatter. To discourage this, the PEEK spacers separating the gate and anode are designed such that adjacent spacers create an S-shaped ``labyrinth" between them (Figure \ref{spacerLabyrinth}). This removes any direct line-of-sight from the outside to the inside of the TPC, reducing the likelihood of light transfer between the two regions. Dark Kapton light baffles located above the anode grid (Figure \ref{ERCutaway}) prevent S1 light ingress from the gaseous xenon regions outside of the TPC. They also reduce inward reflections of S2s near the TPC walls, which substantially improves the quality of the position reconstruction for peripheral interactions, and hence reduces the leakage of wall backgrounds into the TPC's fiducial volume.\footnote{Though not shown in Figure \ref{ERCutaway}, a second set of vertical Kapton baffles was added around the OD of the anode during ER construction to further reduce the chances of light leaks between the skin and TPC.} 

\subsubsection{Liquid level} Precise control over the SE size requires good definition of the position of the liquid xenon surface relative to the extraction region. During normal operation, liquid xenon enters both the TPC and side skin region at the base of the ICV. In the TPC, this liquid rises, then migrates outward near the liquid level through the open PEEK spacer labyrinth. Liquid from both the skin and TPC spills over a set of three weirs (Figure \ref{ERCutaway}) evenly spaced around the TPC. It is then carried out of the ICV into the external circulation path. In this way, the weirs set the liquid level in the detector. These weirs rest on the gate ring, and were machined from PEEK to prevent excessive electric fields on their edges. The labyrinth was designed to have a minimal impedance to flow, so that the liquid height in the TPC is the same as that at the weirs. A set of six equally-spaced capacitive precision level sensors are attached to special PEEK spacers and are able to monitor the liquid level in the detector to a precision of 20 \microns. Together with the weirs, these level sensors are used to ensure that the liquid level stays parallel to the (undeflected) planes of the gate and anode grids.

%% file: 3_GridProduction/GridProduction.tex
\section{Grid Production}
\label{sec:gridProduction}

The grid production process itself needed to be designed so that the final grid parameters met technical requirements. In this section, we discuss the requirements driving the design of the production process. We then describe how the production environment and process were developed to ensure that the final grid parameters met requirements. We then discuss a post-production treatment used to enhance the gate grid's resistance to electron emission. We finish with a discussion of the handling protocols used to prevent the grid parameters from deviating from their requirements between production and installation into the TPC at SURF.

\subsection{Grid Production Requirements}
\label{subsec:EngineeringConstraints}

\begin{figure*}[ht]
\centering
\includegraphics[width=6in]{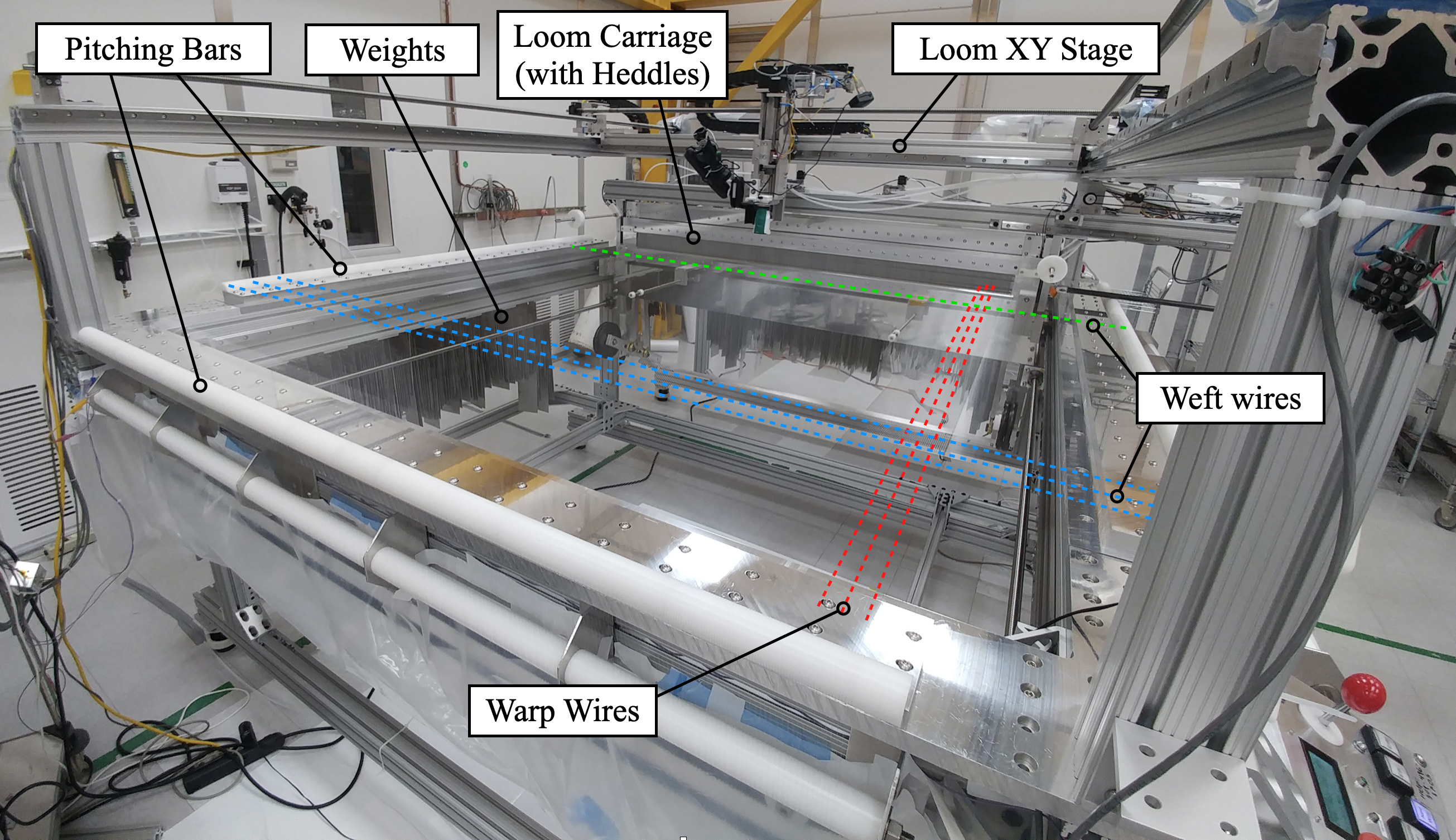}
\caption{A photo of the grid weaving loom and \textit{xy} motion stage. Included are lines indicating the directions of warp (red) and weft (blue) wires along the two axes of the loom. The green line indicates the location of the ``newest" weft wire being woven in relative to the loom's moveable carriage, behind it.}
\label{Loom}
\end{figure*} 

Design of the grid production process was driven by a need for field uniformity, resilient HV performance, and low radiogenic background. 

To achieve a high degree of field uniformity, the grid production process needed to enforce a uniform wire pitch, uniform wire tension, and high degree of precision in flatness and roundness of the rings. These three general constraints were most critical for the extraction region grids because deviations in pitch or tension, or slight non-planarity could directly affect the production of S2 light and limit S2 resolution. 

To optimize the grids' HV performance, every part of the grid production campaign was tailored to limit the introduction of debris and scratches on the rings and wires. Dust and debris is known to increase the likelihood of breakdown, in addition to increasing rates of field electron emission \cite{FNALHVWorkshop}. Likewise, scratches added to the wires or rings can increase local fields and nucleate electrical breakdown.

The effort to limit grid radiogenic backgrounds was largely focused on minimizing exposure to \(^{222}\)Rn. Limiting the exposure of the finished grid wires to plate-out of \(^{222}\)Rn daughters was critical for achieving low rates of \(^{210}\)Pb, \(^{210}\)Bi, and \(^{210}\)Po decays, among which are low-energy betas and a \(^{206}\)Pb nuclear recoil. While many of these grid events can be eliminated from a final WIMP-search dataset by examining the S1-S2 time separation, some are at sufficiently low energies (and occurring in sufficiently strong electric fields) that the S1 may be missed completely. These events can be particularly problematic for ionization-only (S2-only) searches and for accidental coincidence backgrounds in a S1+S2 analysis, because their S2s can occur near the radial center of the detector. If such an S2 piles up with an uncorrelated S1, it may mimic a low-energy WIMP scatter in LZ's search region. It was also important to limit exposure of the grid rings to \(^{222}\)Rn daughter plate-out, primarily because alphas from the \(^{210}\)Po decay can induce neutron emission from \((\alpha,n)\) interactions in the PTFE surrounding the grid rings. This was particularly important for the bottom and cathode grids, whose rings are covered nearly hermetically by PTFE (Figure \ref{TPCDiagram}). Concern over \(^{222}\)Rn emanation from dust on the rings and wires led to the establishment of rigorous cleaning standards for all grid components as well.


\subsection{The Grid Production Environment and Tools}
\label{sec:GridProdEnvironment}

A custom platform was designed to reproducibly construct grids satisfying technical requirements. This platform was built within a class 100 cleanroom at the SLAC National Accelerator Laboratory. The use of a cleanroom not only limited the amount of ambient dust and debris deposited on the grid, but was also advantageous because HEPA airflow is known to reduce the likelihood of \(^{222}\)Rn daughter plateout on the grid wires \cite{tdr}.

\begin{figure}[t!]
\centering
\includegraphics[width=\linewidth]{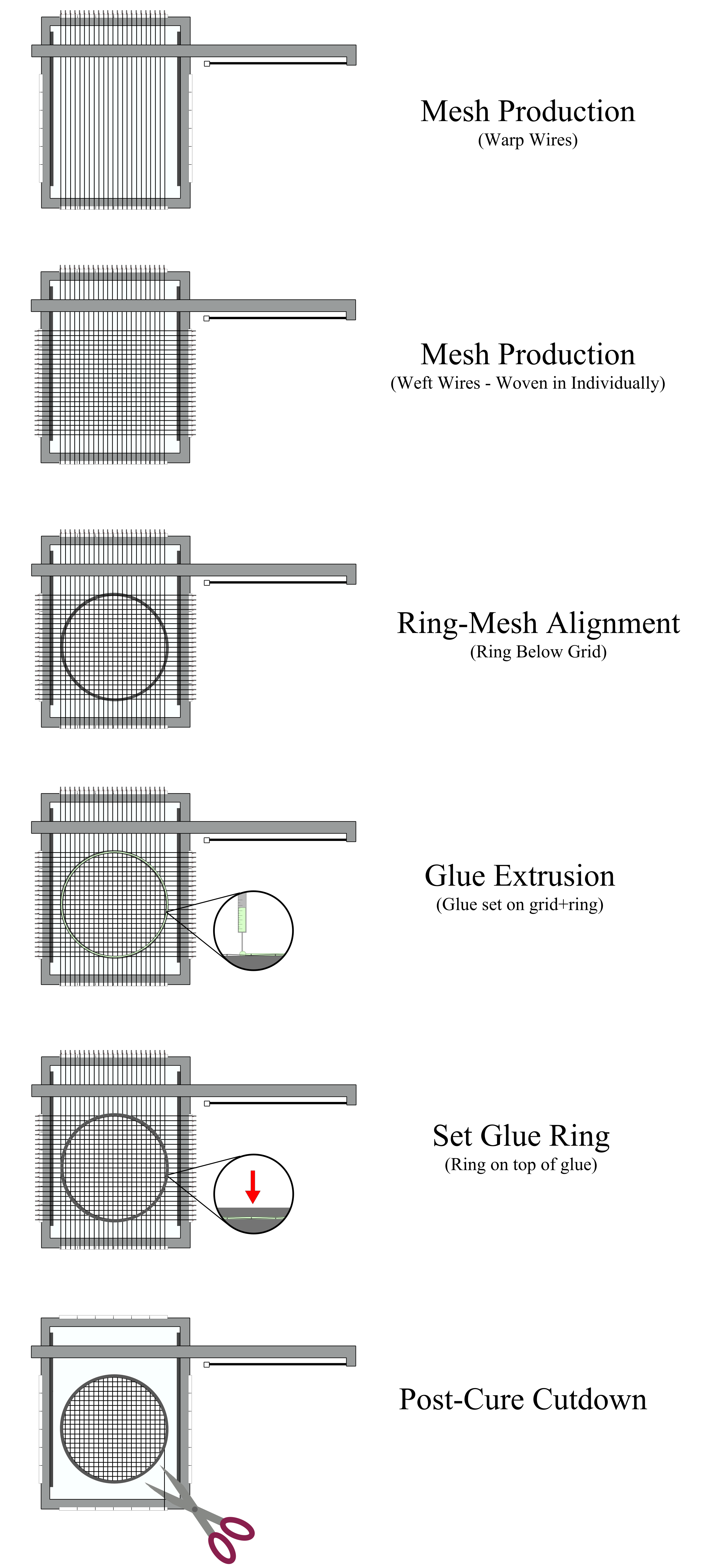}
\caption{A chronology of the steps used to produce a grid, omitting the wire and ring production/preparation. Each image is a top-down diagram of the loom during a particular stage of production. The heddles are located within the moveable carriage, which is represented as the long horizontal bar.}
\label{GridProductionProcess}
\end{figure}

The instrument developed for creating the woven mesh was an 8'x10' mechanical loom (Figure \ref{Loom}). Its design was similar to, and motivated by, looms used in the weaving community throughout history \cite{historicalLooms}. On this loom, a mesh was created in two steps, shown in the top two diagrams of Figure \ref{GridProductionProcess}. In the first, all ``warp" wires were strung along the long axis of the loom, each passing through narrow slots in a pair of vertically movable ``heddles". Then, adjacent warp wires were vertically separated by actuating the heddles, opening a space (the ``shed") through which an individual cross, or ``weft" wire could be inserted. The heddles were again actuated, locking this new weft wire against the warp wires and opening a shed for the next weft wire. This process continued until all weft wires were woven into the mesh.

The additional requirements of consistent tensioning, spacing, and handling of the wires motivated some deviations from the canonical loom design. To achieve uniform wire tension, a 0.25 kg weight constructed from thin stainless steel plate was affixed to the end of each wire and hung under gravity. The relative standard deviation in the weights (and therefore wire tension) was 1.5\(\%\). Each wire end was secured to a plate using an ``envelope button" method, where it was captured between two pieces of silicone rubber and wound multiple times around a brass grommet embedded in the plate until friction prevented it from slipping.


To produce a mesh with uniform pitch, a set of white Delrin plastic ``pitching bars'', visible in Figure \ref{Loom}, were fabricated with v-shaped notches. These notches constrained the wire ends at a pitch of \(4.97 \pm 0.01\) mm, with additional slots halfway between them for the 2.5 mm pitch anode grid. A few inches of extra notches were added to the ends of the pitching bars. This was done so that the mesh could be made larger than necessary, then trimmed to exclude from the final grid any nonuniformity due to mesh edge effects. Maximizing pitch uniformity also required precise alignment of the loom frame and components, which was done through a combination of laser level squares and rigid, custom-machined low-tolerance aluminum braces. Semi-automating the weft insertion process also ensured weft pitch consistency to within 150~\microns\ in the mesh. This jitter is larger than that introduced by the pitching bars, and is due to subtle forces on the wires that were challenging to model and control (see Section \ref{subsubsec:MeshProduction}). The semi-automation also helped minimize the time spent during weaving, resulting in a shorter dust and \(^{222}\)Rn exposure than what would have been achieved with a fully manual process.

To reduce the likelihood of scratching the wires, the heddles were made from polycarbonate. Delrin was chosen for the same reason for the pitching bars; the smooth plastic prevented wire blemish formation as wires were pulled back and forth across the loom during the warp stringing.




A computer-controlled \textit{xy} motion stage was installed above the main loom body to increase reproducibility in the rest of the grid production process. The ``head" of the motion stage accommodated various tools, enabling a variety of different tasks: 
\begin{enumerate}
    \item With a diamond-tipped engraver installed, it provided a reproducible way to score the grid ring surfaces facing the glue joint in preparation for gluing.
    \item With a 3 megapixel Edmund Optics camera (part EO-3112C) and 55 mm telecentric lens, it enabled precision alignment of the ring and wires just before gluing.
    \item With a syringe equipped, it enabled precise dispensing of glue at a pre-programmed rate and height above the mesh.
    \item For the gate and anode grids, it assisted in suspending the glue ring above the grid just after the glue was laid down.
    \item It permitted camera-based inspection (both manual and automated) of the grid for defects and debris. 
\end{enumerate}

\subsection{Production of a Grid}

\begin{figure}[t!]
\centering
\includegraphics[width=\linewidth]{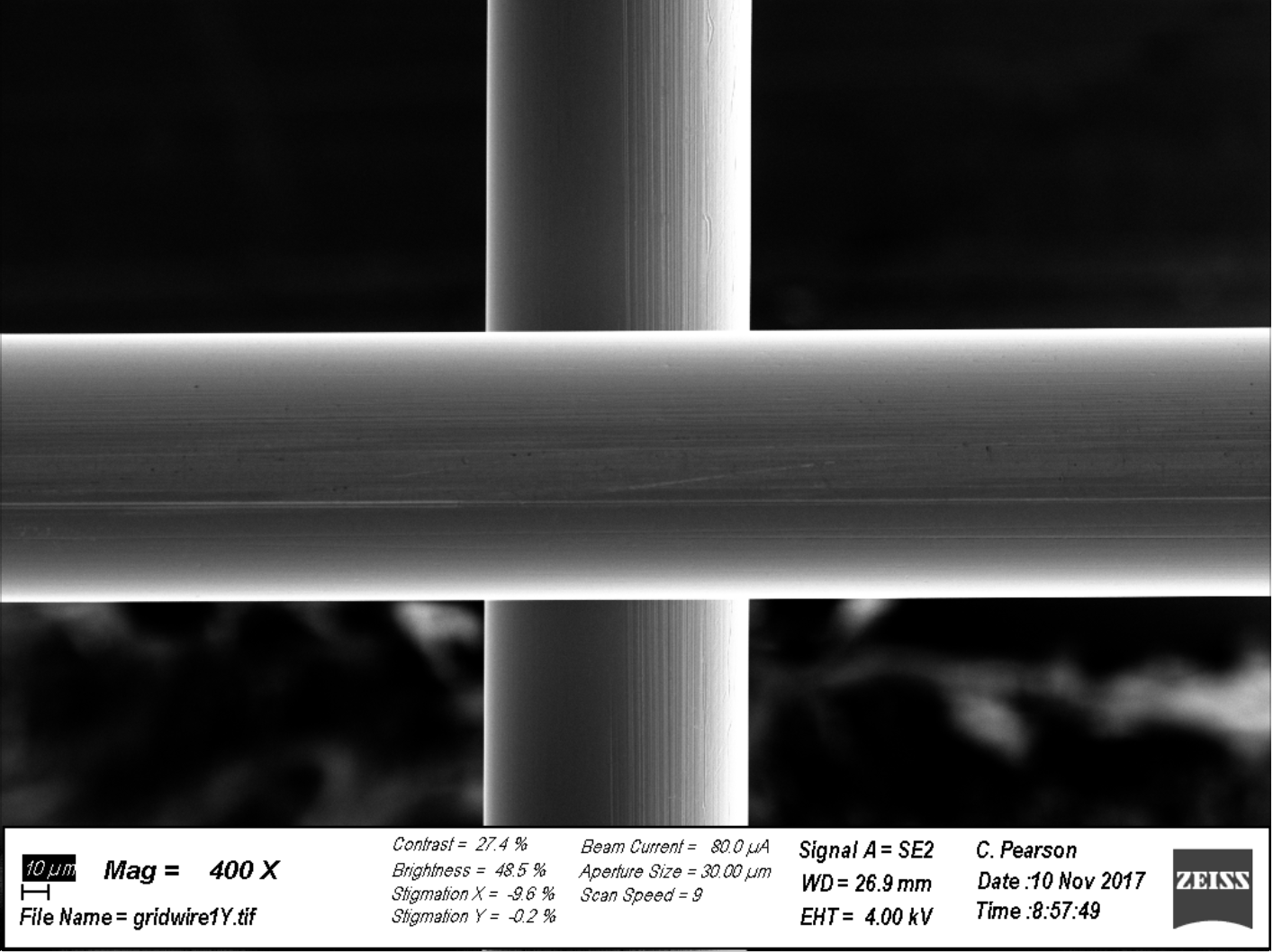}
\caption{A 400\(\times\) magnification SEM image of a cross in a small test mesh woven with LZ wire. Longitudinal striations, presumably from imperfections in the drawing die, can be observed.}
\label{SEMWire}
\end{figure} 

The five sequential elements of the grid production process were:
\begin{enumerate}
    \item Wire production and preparation
    \item Ring production and preparation
    \item Mesh fabrication
    \item Mesh-ring alignment
    \item Grid gluing and cutdown
\end{enumerate} Figure \ref{GridProductionProcess} illustrates the last three steps for visual clarity.

\subsubsection{Wire Production and Preparation} The SS304 grid wire was produced at California Fine Wire and delivered on a set of 10,000-foot spools. At the vendor, the wire was drawn using a dedicated die, ultrasonically cleaned, and inspected at 10x magnification for blemishes prior to spooling. This process was performed within about 8 hours, and the wire was bagged upon completion to limit dust and \(^{222}\)Rn exposure. At SLAC, higher-magnification optical and SEM inspection of wire samples showed that the most common wire blemishes were longitudinal grooves with sub-micron widths, thought to be from microscopic defects in the dies used for drawing (Figure \ref{SEMWire}). Repeated SEM inspections also revealed a wide variety in the appearance of the wire surface, though few samples showed asperities with obvious potential for substantial field enhancement. Similar inspection of wire on a production level would have been useful in building a broader intuition about wire asperities and the quality of the wire surface. However, the logistics and scope of such comprehensive inspection were not conducive to limiting \(^{222}\)Rn exposure or to minimizing risk of creating defects through handling. For this reason, LZ production wire was taken off the spools only during the weaving process.


\subsubsection{Ring Production and Preparation} The cathode and bottom ring sets were machined at the Lawrence Berkeley National Laboratory (LBL) machine shop, and the gate and anode ring sets were produced at Silicon Valley Precision in Livermore, CA. Grid rings were first inspected for deviations from flatness. This was particularly important for the gate and anode, whose glue surfaces were required to be flat to 200 \microns\ to maximize uniformity in the S2 response. 

All grid rings were then electropolished at Advanced Electropolishing in Milpitas, CA. The electropolishing solution used was 65\(\%\) H\(_3\)PO\(_4\), 25\(\%\) H\(_2\)SO\(_4\), and 10\(\%\) deionized (DI) water. The rings were held at +9V relative to the tank walls for 6 minutes. This process was useful not only for its ability to level sharp points which could nucleate electrical breakdown in the TPC, but also for its ability to remove embedded \(^{222}\)Rn daughters, which may be located as far as 100 nm into the surface of the steel~\cite{betaCage}. By removing a few microns of steel from the ring surfaces, electropolishing effectively eliminated \((\alpha,n)\) backgrounds generated by any \({}^{222}\)Rn plateout that occurred since the rings were machined \cite{betaCage}. The rings were also cleaned at Astro Pak, Corp., a precision cleaning company in Los Angeles, CA, to remove any dust, debris, or contamination acquired during or after electropolishing. 

Finally, to increase the surface area for epoxy, a diamond-tip engraver attached to the loom's \textit{xy} stage was used to cut a hatch pattern into each surface facing a glue joint. This hatch also wicked the epoxy away from the radial boundaries of the glue joint, allowing better control over glue distribution prior to adding the glue ring.




\subsubsection{Mesh Production}
\label{subsubsec:MeshProduction}One challenge facing the mesh production was the trade off between rigorously examining the wires for defects and limiting the wires' exposure time to \(^{222}\)Rn and dust. Even though grid production was done in a cleanroom where HEPA flow may have reduced radon plateout, it still represented a non-negligible fraction of the open-air exposure of the wires to \(^{222}\)Rn. As a result, a quick visual inspection was performed on each wire to ensure that it was free of kinks and macroscopic blemishes, and to monitor the uniformity of the weave as each wire was added. This inspection generally added 10-20\(\%\) to the insertion time for each wire, giving a production time of three days for an average 5 mm pitch mesh.

\begin{figure}[t!]
\centering
\includegraphics[width=\linewidth]{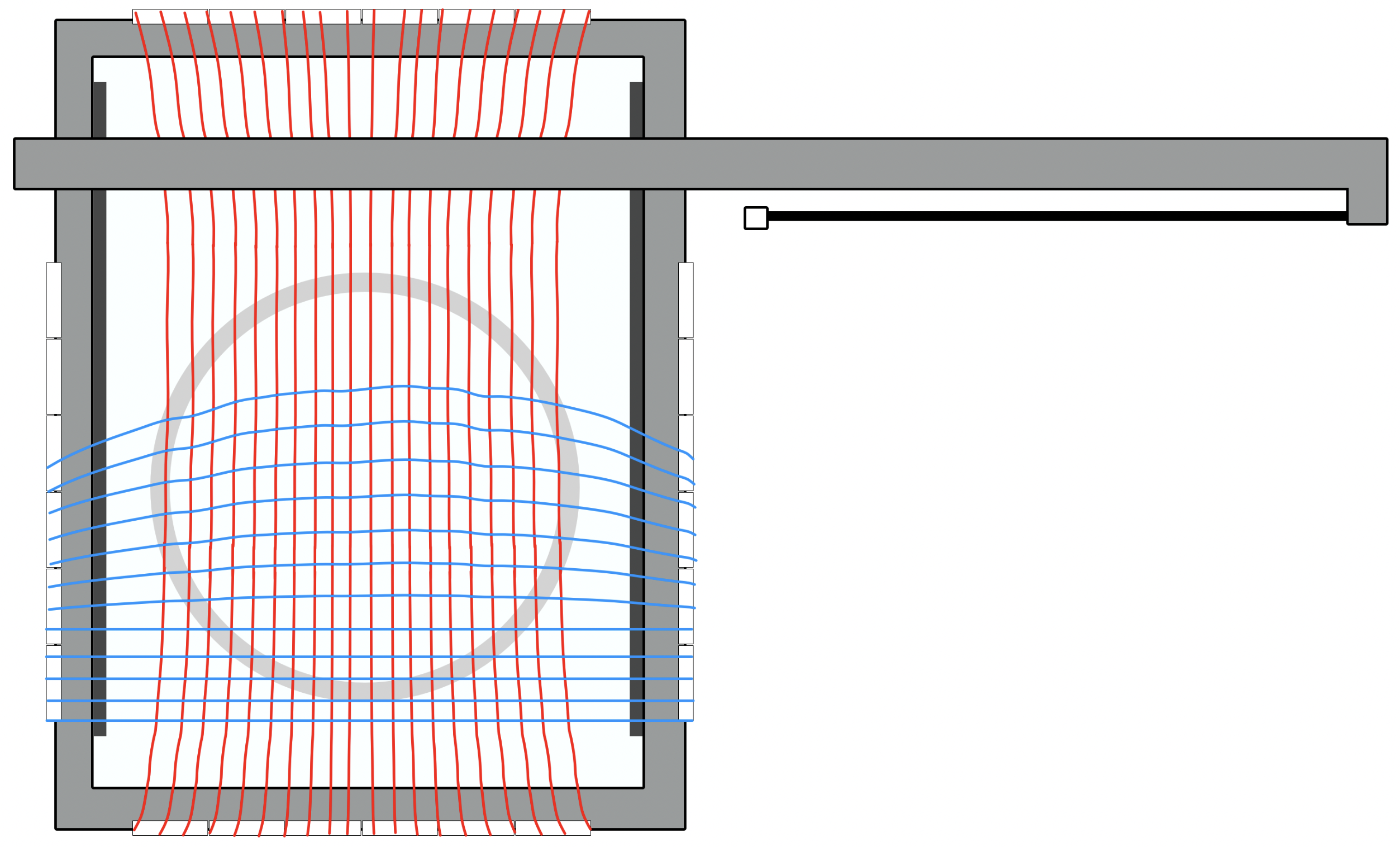}
\caption{The two kinds of weave distortions observed while producing a mesh. A grid ring is superposed for reference. The scale of distortion has been significantly exaggerated for visual clarity. The red (warp) wires tapered inward relative to their pitching bar positions. The blue (weft) wires were pushed by the shed warp wires to create an outward ``bow" that became more prominent further into the grid. Both kinds of distortion were more prominent on the anode weave.}
\label{LoomWireDistortions}
\end{figure}

While producing the tighter, 2.5 mm pitch anode grid, two additional obstacles were encountered, both related to distortions in the uniformity of the weave. The first was a ``bowing" of new weft wires away from the existing mesh due to contact with the warp wires defining the shed, as shown in the blue lines in Figure \ref{LoomWireDistortions}. This was eliminated by introducing a comb-like loom component (the ``reed'') to better enforce uniform weft spacing. Warp wires passed through a set of reed ``teeth,'' which held each new weft wire in a correct position as it was woven into the grid. The second obstacle was a global inward taper of the warp wires, as shown in the red lines in Figure \ref{LoomWireDistortions}. The most extreme taper was measured to be a difference of 0.4 mm in a warp wire's position from the front to the back of the full square mesh (1.0 mm from the front pitching bars to the back of the mesh). This taper was a result of the warp wires being pulled inward by friction from each weft wire as it was woven into the grid, and there was no simple way to fix this.

\begin{figure*}[t!]
\centering
\includegraphics[width=\linewidth]{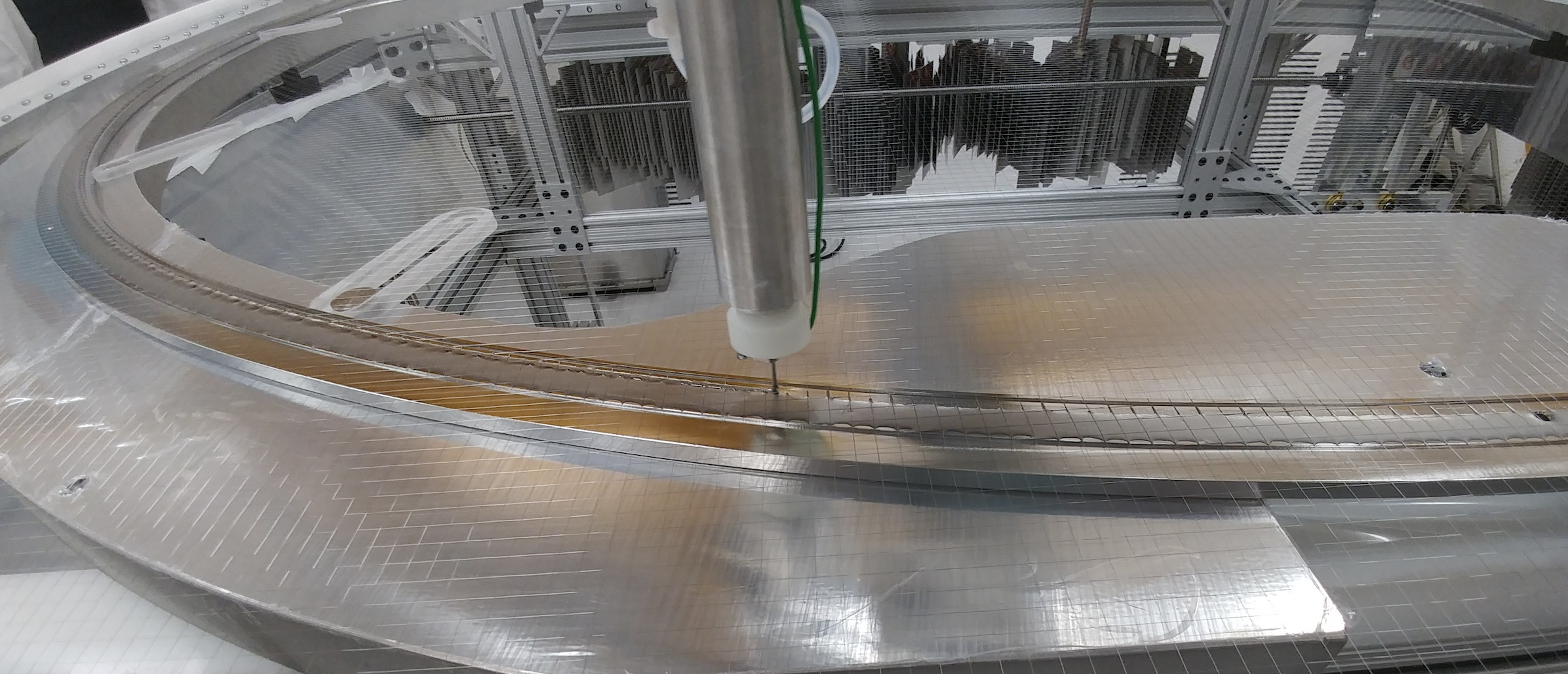}
\caption{A photo of the loom \textit{xy} stage extruding epoxy during the bottom grid gluing.}
\label{Gluing}
\end{figure*}

Similar to the alignment of the gate and anode, jitter in wire pitch is expected to cause some amount of S2 smearing, though the scale of smearing has not been studied exhaustively. However, for the purpose of informing the fabrication of future large-scale electrodes, we find it worth noting here which effects are thought to have contributed most to nonuniformities in the weft and warp wire spacing. Due to the sparser 5 mm weave, nonuniformity in the gate weft spacing was likely driven by the precision in the pitch of the notches of the pitching bar. The gate warp wires had some inward taper similar to that of the anode, but the scale of this taper was not explicitly measured, and was expected to be negligible. The anode warp wire pitch distribution was largely governed by the reed, whose teeth contacted the warp wires and limited the inward taper for much of the grid. The resulting pitch variations were typically on the scale of 100 microns. The anode weft wire spacing was governed by the precision with which the reed could constrain a new weft wire at exactly 2.5 mm from the last. The reed was used in conjunction with mounted-camera-based inspections of the wires to achieve anode weft spacing consistent to within 150 \microns\ of the desired 2.5 mm pitch.



\subsubsection{Mesh-Ring Alignment} Having alignment between the gate and anode meshes was motivated by considerations of S2 resolution (Section \ref{subsubsec:S2Design}). As a result, an alignment step was developed to properly position grid rings relative to their corresponding meshes before gluing. In this step, a ring was set on a support platform under the mesh and carefully adjusted until a set of four vertical dowel pin holes machined in the ring were aligned with the squares of the mesh. The ring was then raised up against the mesh, using the loom \textit{xy} stage's camera to ensure that the mesh and pin holes remained aligned throughout the process.

The metric for successful alignment between a mesh and its ring varied from grid to grid. The most important target for alignment was a match between the anode and gate grid wires as shown in Figure \ref{Alignment}. 
This process was challenging, in large part because of the taper shown in Figure \ref{LoomWireDistortions}. As a result, the full extraction region could not be aligned perfectly. With help of the loom \textit{xy} stage's camera and through careful, 50-\microns-scale adjustments of the rings, the central regions of the gate and anode grids were precisely aligned at the cost of losing alignment in the outer regions.

Because the cathode and bottom grids are positioned much lower in the TPC, the detailed field structure near the wires has no impact on S2 resolution. As a result, there was no requirement on their relative alignment.

\subsubsection{Mesh-Ring Gluing and Cutdown} \label{subsubsec:Gluing} The effort required to create a precise, robust, and reproducible glue joint made gluing the most technically challenging part of production. A syringe mounted on the automated loom \textit{xy} stage dispensed a continuous bead of glue around the ring at a height of 300 \microns\ above the wires (Figure \ref{Gluing}). Millimeter-scale non-planarity and eccentricity in the flexible bottom and cathode rings required height and \textit{xy} corrections to this circular path. After this, the PMMA beads used to set the glue joint thickness were deposited in 750 nL glue droplets at 100 evenly-spaced points around the ring. The total volume of glue was selected to overfill the expected width, height, and circumference of the glue joint by 10\(\%\) to ensure that epoxy effectively wetted both the glue and base ring surfaces facing the joint. After glue extrusion, the glue ring was placed onto the wires, using the alignment dowel pins to guide the ring placement. Ensuring the absence of gaps between the two rings was critical to the integrity of the final grid. Because all of the glue rings were quite flexible and did not lie flat on the base rings, it was necessary to apply downward pressure on them to close gaps in the glue joint. This was done with T-slot bars for the bottom and cathode, and a set of spring-loaded plungers mounted to a frame for the particularly delicate gate and anode glue rings.


Another major challenge encountered in the gluing process was the development of glue cure method that both limited \(^{222}\)Rn exposure and ensured a strong glue bond. The recommended epoxy cure schedule was for a short time at elevated temperature, but the manufacturer suggested that a week long room temperature cure was acceptable. Tests with small, 15 cm diameter grids demonstrated that short, day-scale heat cures at 72\(\degree\)C generally produced a stronger glue bond than week-long cures at room temperature. However, at the 1.5 meter scale, it was impractical to heat the full grid assembly, and heating the ring alone would cause wire detensioning upon cooling. For a 70 \(\degree\)C cure temperature, such a ring-only heating would give a \(55\%\) loss in anode wire tension. As a result, a room temperature cure between 5 and 15 days was used for the LZ grids, with the variation driven by overall project schedule. This room-temperature cure schedule was also used for several 15 cm diameter prototype grids to test if the glue joint would withstand thermal cycling. These tests confirmed the cryogenic robustness of glue joints made with this cure schedule.

Once the glue was cured, the grid was cut out of the mesh just on the external diameter of the glue joint. To complete the grid production process, the guard ring was bolted in (for the cathode and bottom) or laid (in the case of the gate and anode) into place on the base ring.





\subsection{Grid Passivation}
\label{subsec:Passivation}
In addition to limiting the creation of field-enhancing asperities on the wires during production, it was also possible to actively reduce the emission from any existing asperities or imperfections in the oxidized metal surface through a chemical process called passivation.

\subsubsection{Motivation for Passivation}
It has been shown in a recent R\(\&\)D study in the context of LZ that passivation following acid cleaning reduces field emission from electrode wires similar to those used in LZ \cite{ImperialWireStudies, SLACcitricpaper}. This may stem from the passivation process altering the surface chemistry of cathodic electrodes to remove corrosive elements from the surface and rebuild a more corrosion-resistant electrode surface  \cite{OxideNoer,OxideStygar}.  Processes that remove ferric oxides from stainless steel surfaces and thus increase the chromium-to-iron ratio include pickling or acid cleaning to strip the stainless steel's existing oxide layer. These processes typically involve treating a degreased stainless steel surface with a solution containing nitric acid and often at elevated temperatures. 
After acid cleaning, a new oxide layer will form in 24--48 hours to complete the passivation process. This rebuilding of an oxide layer in very controlled conditions is central for the reduction of field emission as demonstrated in silicon electrodes~\cite{OxideHuang,OxideYang} and supported by studies on stainless steel wires~\cite{ImperialWireStudies}. 

To determine the optimal passivation treatment for LZ, small prototype grids constructed with the design wire and mesh specifications were passivated using either a room temperature nitric acid solution or a heated citric acid solution, and electron emission from these grids was measured in small xenon testing platforms at SLAC~\cite{kitchenSinkST}. These studies concluded that the citric acid passivation process was highly effective in reducing the rate of electron emission from cathodic grids \cite{SLACcitricpaper}. Citric acid was therefore selected for use on the LZ gate grid, with the added benefit that it is considerably easier to handle than nitric acid in large quantities.

\subsubsection{Passivating the Gate Grid} 
Because the gate grid is a cathodic electrode in the electron extraction region and has the highest surface fields of any grid, it is the most likely to experience field-induced electron emission which may lead to accidental coincidence events. LZ's gate grid was passivated to reduce this susceptibility to electron emission.    

The gate grid was passivated in a cleanroom at Astro Pak, Corp. in a proprietary solution comprised of 3--5\% citric acid for two hours at a temperature of $(124.8 \pm 2.0)^{\circ}$F. After submersion in the citric acid bath, the gate grid was thoroughly rinsed with deionized water and stored in a custom bag backfilled with filtered air to rebuild the chromium-enriched oxide layer in a dry environment while protecting the gate grid from dust exposure. 

It is worth noting that during prototype grid testing, heated passivation baths weakened room-temperature cured EP29LPSP glue joints, but not glue joints that had experienced high-temperature cures. For this reason, the glue joint of the LZ gate grid was strengthened after the grid was taken off the loom, but prior to passivation with a post-cure process, in which it was held at 110\(\degree\)F for 48 hours, then held at 130\(\degree\)F for 24 hours.

\subsection[Grid Handling, Cleanliness, and Radon Exposure]{Grid Handling, Cleanliness, and\\ Radon Exposure}

\begin{figure*}[t!]
\centering
\includegraphics[width=\linewidth]{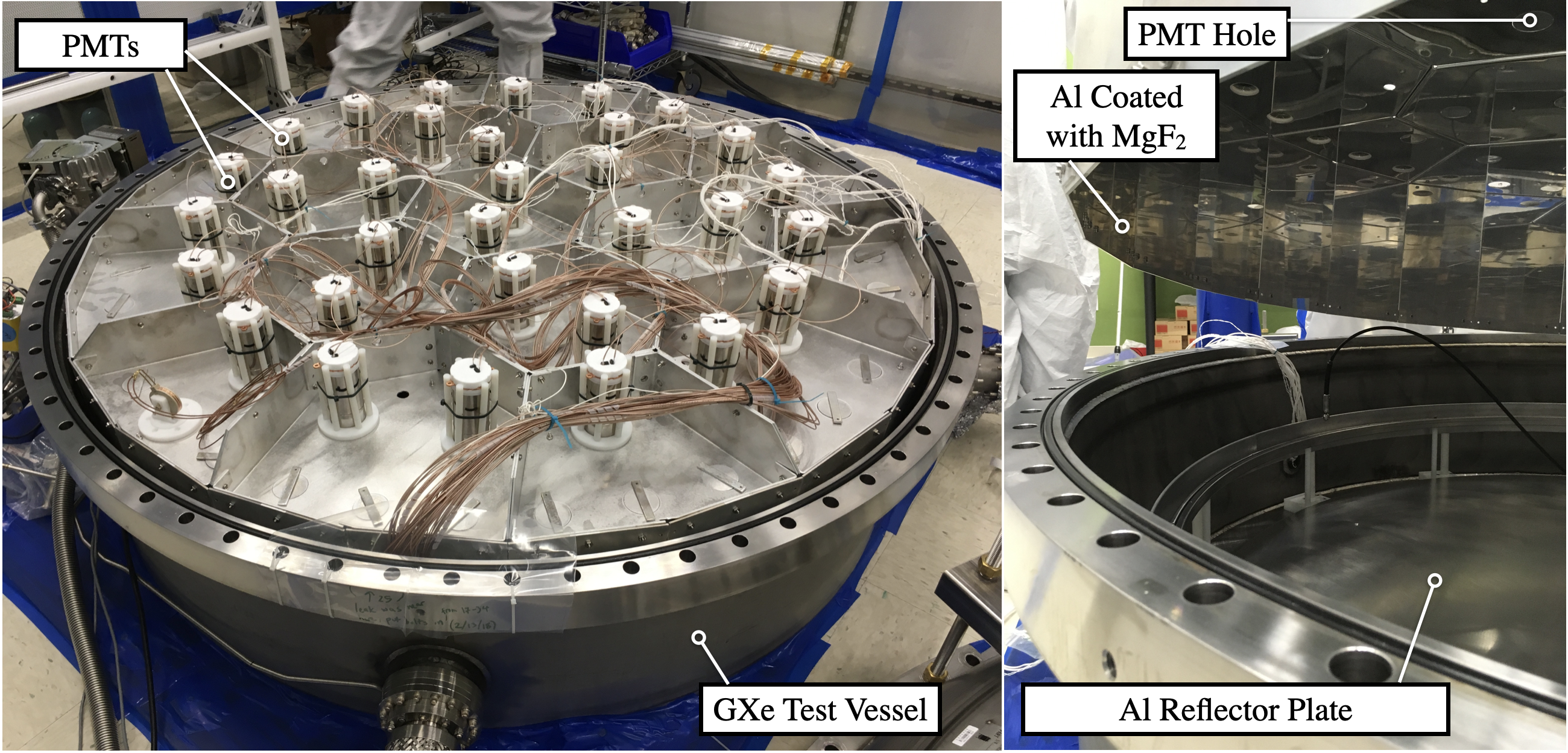}
\caption{Photos of the gaseous xenon test chamber used for HV performance testing of the full-size extraction region and grids. Left: the chamber with the PMT array installed. Right: the chamber with the array being lowered, showing the reflective aluminum inner surface and lower reflector plate.}
\label{fig:PhaseII}
\end{figure*}

To optimize HV performance and minimize radiogenic backgrounds, it was critical to keep the grids clean between their construction and their installation into the TPC. However, due to the 1.5 meter scale, it was impossible to perfectly protect them from dust, even in a cleanroom. Ultraviolet light (365~nm) was used to induce fluorescence in dust, making it easier to see, but a complete manual examination of the wires for dust in this manner was impractical. Dust shed from cleanroom suits also made it possible for someone to add more debris faster than they could manually identify and remove it. A method was developed with the loom \textit{xy} stage to exhaustively scan the grid for dust under UV lighting. This method was effective at quantifying and characterizing the amount and location of debris on the grid, and helped build an understanding of the level of cleanliness achieved during grid operations such as passivation. Spray washing a grid with deionized (DI) water and drying with filtered N\(_{2}\) was found to be highly effective at removing dust and debris of all sizes visible with a UV lamp. Such a spray, coupled with a final, rapid, visual UV inspection of the grid, was used at SURF immediately prior to grid installation.

Protection of the rings and wires from \(^{222}\)Rn also drove many handling protocols. Wire spools were kept in an N\(_{2}\) purge box when not in use. Rings were tightly double-bagged in nylon, with an N\(_{2}\) backfill to limit the exposure from ambient-air radon when possible. Completed grids were double-bagged with nylon and backfilled for the same reason. Because early attempts at this process revealed that wrinkles in the bags could displace grid wires, a separate taut ``drumhead" of polyethylene film was placed on either side of the grid within the nylon bags to eliminate bag-to-wire contact. Installation of the grids into the TPC at SURF was done in a Rn-reduced cleanroom environment, keeping the wires' and rings' exposures low during this process~\cite{cleanlinessPaper}.




%% file: 4_GridERValidation/GridERValidation.tex
\section{Grid and Extraction Region Validation}
\label{sec:validation}

After grid production, a set of measurements and tests were made to confirm that the grids and extraction region would reach their design performance in LZ. These included tests of HV performance, measurements of the gate-anode alignment, measurements of the gate-anode deflection, and \emph{a priori} estimates of the radiogenic background levels of the four grids, with an emphasis on those from \(^{222}\)Rn exposure.

\begin{figure}[b!]
\centering
\includegraphics[width=\linewidth]{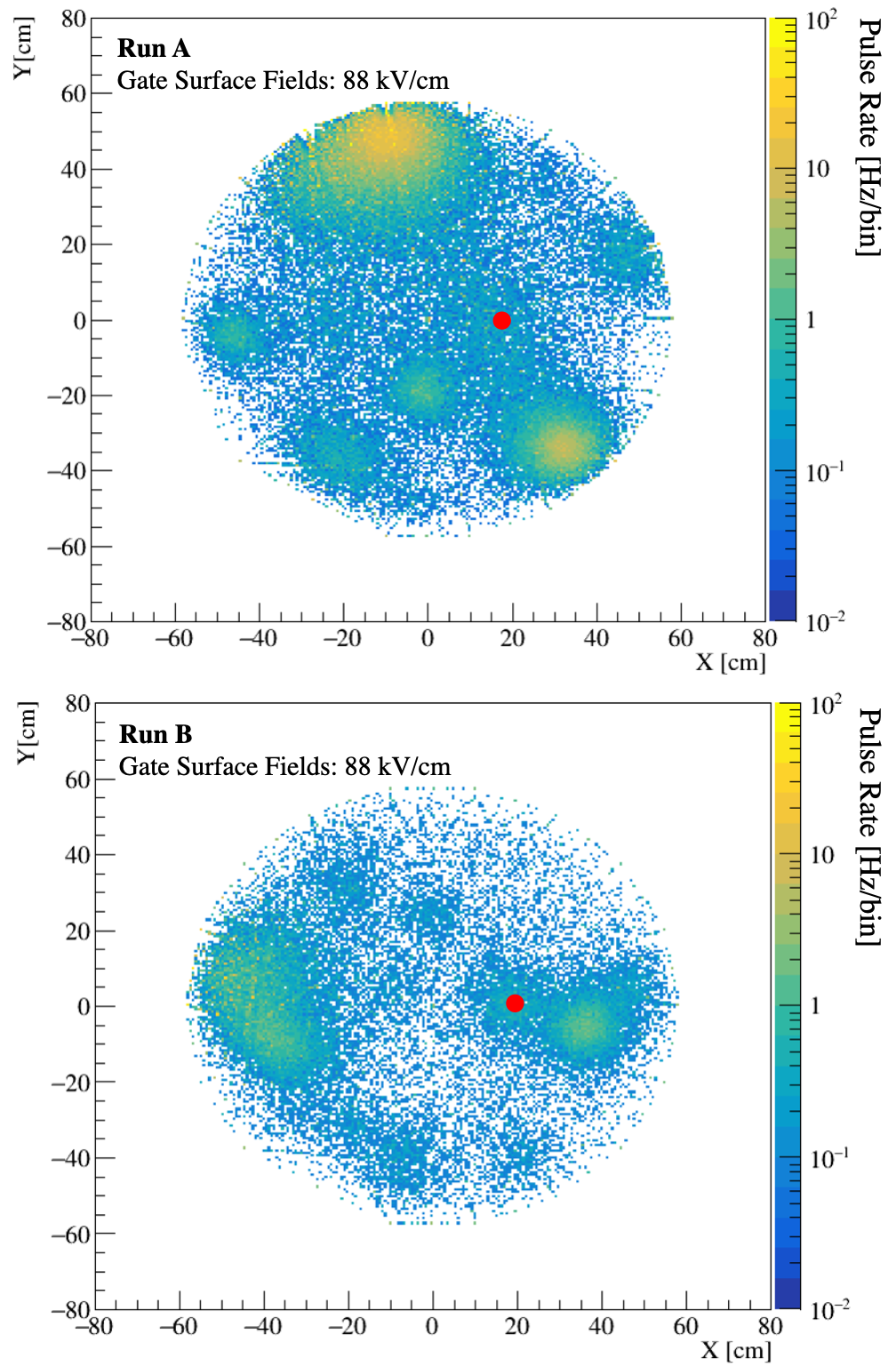}
\caption{Rates of SE-like pulses observed from the gate in the SLAC extraction region GXe tests, which were done prior to passivating the gate. The position of a pulse was built from the \textit{xy} centroid of the photoelectrons within the pulse. In Run A (top), several localized sources of SE-like pulses could be observed when the ER was taken to a \(\Delta V\) of 12 kV. Between Run A and Run B (bottom), the chamber was opened and some dust was manually removed, but some new dust was also added to different parts of the grid. Run B shows a different set of hotspots, suggesting a link between transient debris and electron emission. Note that in Run A, there are a few bins for the topmost hotspot where the rate exceeds 100 Hz. However, this \textit{z}-axis range was chosen to provide sufficient dynamic range to compare this run to Run B, where such high rates were generally not observed. A red dot is used to indicate the hotspot at (19,0) which persisted through every run taken with the ER.}
\label{fig:Hotspots}
\end{figure}

\subsection{High Voltage Performance} 

The gate and anode were assembled with their PEEK spacers within a dedicated GXe pressure vessel in a cleanroom at SLAC (Figure \ref{fig:PhaseII}). An enclosure supporting an array of 32 PMTs was placed above the assembled extraction region. The enclosure walls were made from aluminum and were coated with MgF\(_{2}\) to limit oxidation and maintain 90\(\%\) reflectivity to Xe scintillation light. A reflective aluminum plate, also with MgF\(_{2}\) coating, was placed below the grids. The chamber was filled with room-temperature xenon to 3.3~bar(a), which gives a density equivalent to that of cold xenon gas in LZ. Gas was periodically circulated through the chamber and through a getter to eliminate electronegative impurities.

Two sets of tests were performed using this chamber and its gas environment: one studying the breakdown voltage of the extraction region, and one studying rates of electron emission from the gate. Both sets were done prior to passivating the gate grid. In the former, the PMTs were turned off. The \(\Delta V\) between the gate and anode was increased until electrical breakdown occurred at 10~kV. Follow-up tests achieved marginally higher voltages before sparking, showing stable performance at 11~kV. Electrical breakdown predominantly occurred along the surface of the PEEK spacers separating the two rings. The breakdown was attributed to two causes. The first was a triple junction point field enhancement caused by the presence of a chamfer on the edges of the PEEK spacers near the gate and anode grid rings. A new set of spacers without a chamfer was machined for use with the final extraction region. The second cause of breakdown was the lack of a field-suppressing liquid layer covering the gate. For a design gate-anode \(\Delta V\) of 11.5~kV, gate surface fields are \(60\%\) higher in xenon gas than they are with a liquid level 5~mm above the gate. The much shorter electron mean free path in liquid suppresses electroluminescence and Townsend avalanche that could otherwise occur due to potential high-field regions at the PEEK-gate interface. This further increases the breakdown \(\Delta V\).\footnote{Evidence for this effect was demonstrated in a separate small-scale test platform at LBL~\cite{SorensenLBL} in 2017.} By reaching and holding a \(\Delta V\) of 11~kV, these tests confirmed that the ER could hold gate surface fields above the LZ design voltage. Though this \(\Delta V\) only probed anode surface fields up to approximately 80\(\%\) of the design fields in LZ, the modified PEEK spacers and liquid level are expected to increase the gate-to-anode breakdown voltage to the design value of 11.5~kV.

In tests of electron emission, the PMTs were turned on. Together with the reflective enclosure, the PMTs gave the chamber an expected LCE of about 1\(\%\), enabling some sensitivity to single electron emission from the grids. The PMT array also allowed for some coarse localization of electron emission in \textit{xy}. For these tests, all but 6 PEEK spacers were removed to reduce stressed area near the PEEK chamfer and increase the maximum \(\Delta V\) that could be held without electrical breakdown. In each test, the \(\Delta V\) between the gate and anode grids was stepped up to 12~kV in intervals of 1~kV, and at each step a rate of electron emission was measured over the span of several seconds. These tests showed a variety of spatially separated emission hotspots across several data-taking periods, in between which the chamber was opened, inspected, and re-closed. Some of these hotspots either appeared or disappeared from run to run, and were attributed to the addition or removal of small debris (Figure \ref{fig:Hotspots}). One, located at X=19~cm and Y=0~cm in Figure \ref{fig:Hotspots}, persisted through multiple runs and was attributed to a permanent wire defect. Even though the LZ gate was not tested again after passivation, separate dedicated studies on 15 cm diameter grids showed strong evidence that passivation successfully reduces electron emission from features similarly thought to be permanent wire defects. A detailed account of these passivation tests and their results will be given in a forthcoming publication~\cite{SLACcitricpaper}. In light of these studies, the tests on the full extraction region highlighted the importance of both cleanliness and passivation in reducing electron emission from cathodic grids.\footnote{Tests of electron emission from the cathode showed qualitatively similar results.} They further motivated passivation of the gate and strongly motivated the final water spray wash of the grids before installation into the TPC at SURF.

The gate and anode HV cables and cable connections were also tested to ensure that they reached design performance. The gate connection was tested in room temperature, circulating, 3~bar(a) xenon gas, and and was held for prolonged periods of time at voltages between \(-\)7 and \(-\)8~kV. It was also tested in liquid, where a 15-cm scale gate-anode system held a \(\Delta V\) of 11.5~kV for several hours. These two tests demonstrated the high voltage robustness of the LZ gate cable termination. An additional test was done to validate the performance of the final anode termination as it is positioned relative to the PMT array and ICV wall. This test was done in room temperature, non-circulating, 3.2~bar(a) xenon gas, and demonstrated that a short section of the anode ring design could be brought to 8~kV and held for extended periods of time without sparking to either the wall or the PMT array.

\begin{figure}[h!]
\centering
\includegraphics[width=\linewidth]{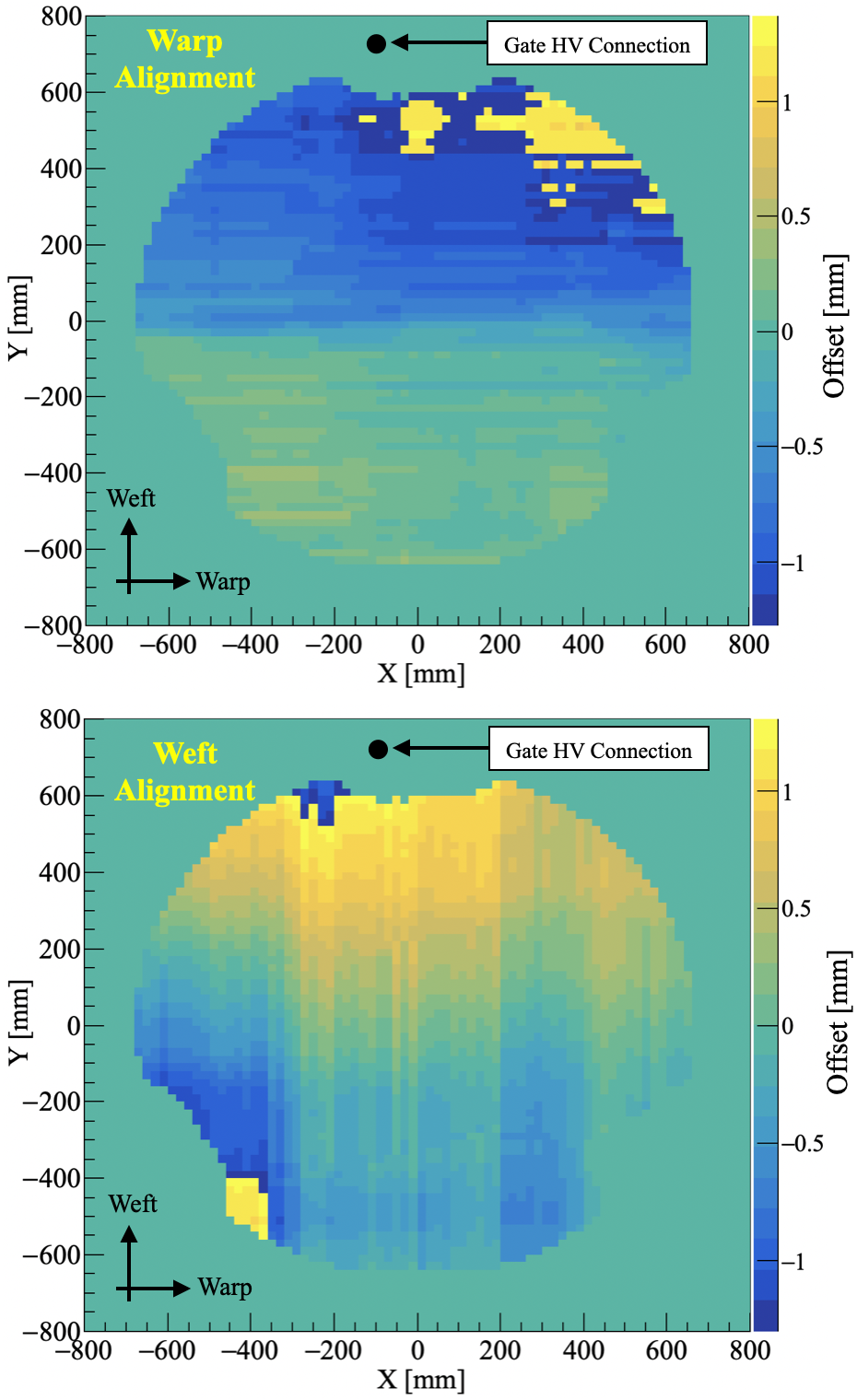}
\caption{Measured alignment of the gate and anode grids, using photos taken from the loom \textit{xy} stage camera. The top plot shows the distance between a gate warp wire and its nearest anode warp wire. The bottom plot is the same, but for weft wires. The regions missing from the top, bottom left, and bottom right of the circular grid are due to the shape of the support platform used to hold the ER during the photo-taking scan, which made it difficult to take pictures useful for assessing alignment in these areas. The regions of the plot where dark blue transitions to bright yellow are an artifact of the use of the ``nearest anode wire" metric in matching anode and gate wires, and do not represent discontinuous transitions in grid pitch.}
\label{AlignmentClocked}
\end{figure}

\subsection{Gate-Anode Alignment} Figure \ref{AlignmentClocked} displays the alignment between the gate and anode grids throughout the extraction region, as measured post-grid-production by the loom \textit{xy} stage camera. In this measurement, the camera rastered across the completed extraction region and at each point in \textit{xy} took one image focused on the gate and one image focused on the anode. These images were then compared to identify the warp and weft wire offsets at that point. Near the center of the grid, warp and weft alignment between gate and anode was achieved to within 100--200~\microns, but some outer regions of the grid achieved a near perfect anti-alignment of the two meshes, where there was a 1.25~mm distance between the closest gate and anode wires. From the discussion of S2 resolution in Section \ref{subsubsec:S2Design}, this implies that the grid alignment contribution to LZ's S2 resolution will be  0.02\(\%\) at 2.458~MeV in the center of the detector, degrading to around 0.1\(\%\) near the edge.

\subsection{Gate-Anode Deflection} 
\begin{figure}[t!]
\centering
\includegraphics[width=\linewidth]{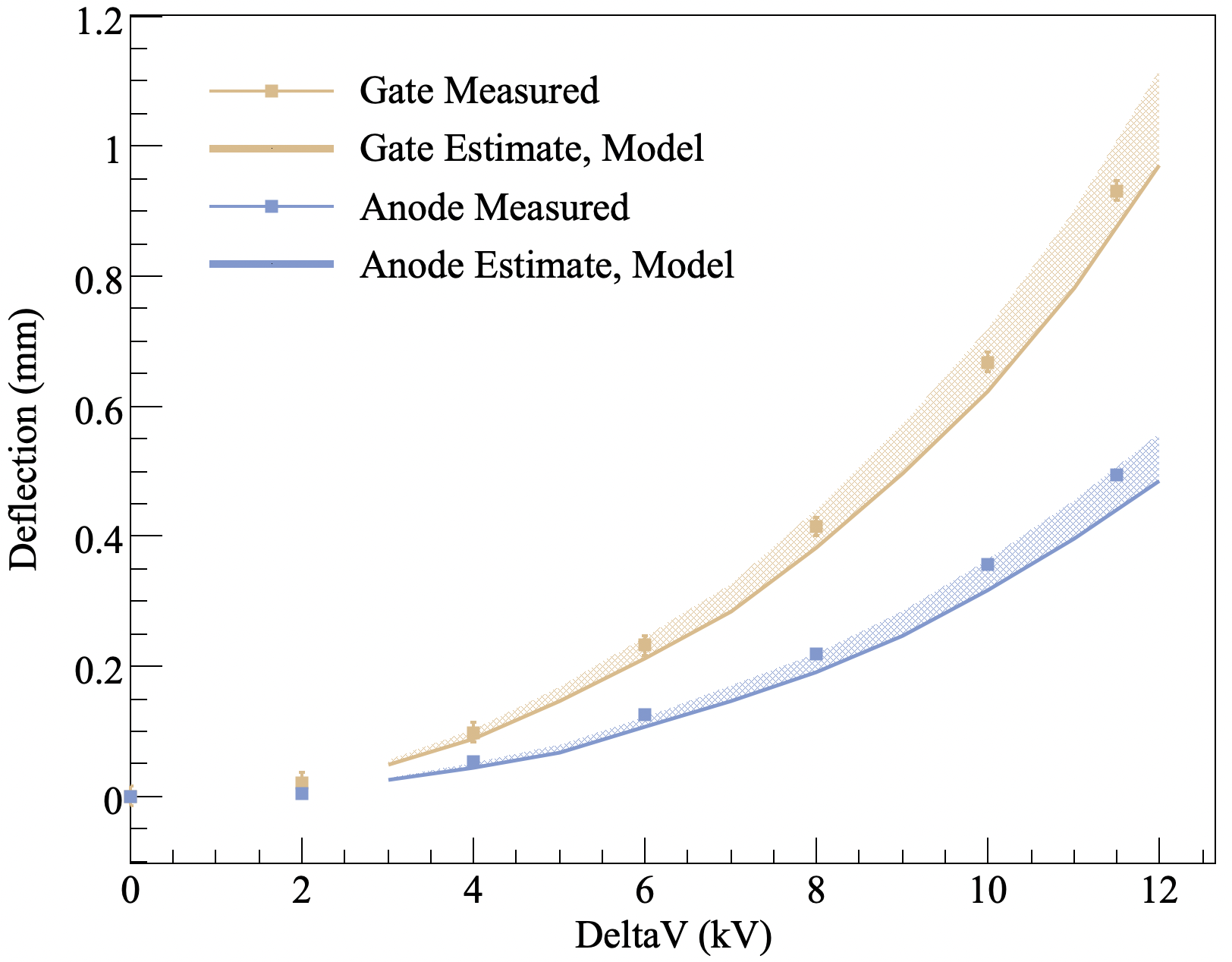}
\caption{\label{fig:dtest}Measured deflections of the gate and anode in air, as a function of \(\Delta V\) between them. These data are compared with the models used to predict deflection in LZ, adjusted to an air-only scenario. The shaded region indicates a model uncertainty associated with the impact of the PMMA beads and cure time on the strength of the glue joint. The uncertainties on the data points are shown, and most are approximately the size of the data point.}
\label{MeasuredDeflections}
\end{figure}

The absolute deflections of the gate and anode in the assembled ER were measured in air in the cleanroom at SLAC using a contactless optical method. At each \(\Delta V\), an optical camera attached to a precision linear actuator was moved until the wires appeared in focus, and the camera's position relative to the zero-deflection position gave the absolute wire deflection. The results of this test are plotted in Figure \ref{MeasuredDeflections}. The model used to predict the grid deflection in LZ for the 2.45~N wire tension was modified to account for a gas-only environment and is consistent with the measured values. Using this validated model, the sum of the expected gate and anode deflections during nominal LZ operation (anode/gate =\(\pm\)5.75 kV) is 1.8~mm, which is within the 2~mm requirement. This accounts for a 25\(\%\) increase from the baseline model (the model corresponding to the solid lines in Figure \ref{MeasuredDeflections}) due to slight weakening of the glue bond from variable cure times, introduction of the PMMA beads, and in the case of the gate, the passivation process.

\subsection{Estimates of Radiogenic Background}\emph{A priori} estimates of the background due to \(^{222}\)Rn daughter plateout are challenging, primarily due to the systematic uncertainties in the fraction of daughters that attach to the wire surfaces. A model of this background has been made using estimates of the ambient radon concentration, exposure times of grid components, air column heights during these exposure times, and the radon daughter plate-out fraction.\footnote{This corresponds to the ``Borexino'' Model in section 12.4 of Ref. \cite{tdr}} Optimistic and pessimistic estimates have been made, with the largest difference between them caused by the uncertainty in how much reduction in plateout is achieved in the presence of a HEPA cleanroom flow. Due to this uncertainty, these estimates are especially sensitive to the details of how long each grid spent in the production and testing cleanrooms at SLAC, which varied substantially from grid to grid. The two estimates are shown in Table \ref{table:Rn222}. While the true rate of these decays can be further constrained through in situ data analysis, preliminary studies with these \emph{a priori} estimates have suggested that achieving rates closer to the optimistic estimate is one condition for being able to discern \(^{8}\)B events above this grid background in an S2-only analysis.

The radiogenic background expected due to the wire's intrinsic radioactive contaminants was measured by assaying a separate spool of wire from the same production batch as that used for the final grid wire \cite{cleanlinessPaper}. Only upper limits on \(^{238}\)U\(_{e}\), \(^{238}\)U\(_{l}\), \(^{232}\)Th\(_{e}\), \(^{232}\)Th\(_{l}\), and \(^{40}\)K could be measured, and those limits give activity levels similar to the optimistic estimate of background from \(^{222}\)Rn daughter plateout.

\begin{table*}[htpb]
\centering
\caption{Estimates of the cumulative time that the wires were open to air in the SLAC production and testing cleanrooms (CRs). In the former, the HEPA flow was oriented normal to the grid plane, and in the latter, the HEPA flow was oriented parallel to the grid plane. The wire from all grids was drawn in a non-cleanroom environment, adding 8 hours onto each of these values. The sum of the times shown and the drawing time represents the majority of the radon exposure for the wires. High and low estimates of the \(^{210}\)Pb activity on the wires due to the total estimated exposure are also provided. Decays from \(^{210}\)Pb and from its daughters \(^{210}\)Bi and \(^{210}\)Po are expected to be approximately in secular equilibrium by the time LZ begins taking data.}
\begin{tabular}{|p{1.5cm}||p{3.2cm}|p{3.1cm}|p{2.9cm}|p{2.9cm}|}
\hline
Grid & Exposure time, production CR (hrs) & Exposure time, test CR (hrs) & \(^{210}\)Pb Activity (Low) (mBq) & \(^{210}\)Pb Activity (High) (mBq)  \\
\hline
\hline
Anode & 1032 & 312 & 1.4 & 88.0\\
Gate & 528 & 456 & 0.8 & 27.4 \\
Cathode & 360 & 24 & 1.2 & 16.3 \\
Bottom & 960 & 0 & 0.3 & 21.6 \\
\hline
\end{tabular}
\label{table:Rn222}
\end{table*}
\vspace{5mm}

%% file: 7_Conclusions/Conclusions.tex
\section{Conclusions}
\label{sec:conclusions}

The four LZ high voltage electrode grids were designed and built in a program from 2014 to 2019. The primary goals of the design and production processes were to optimize field uniformity and electrostatic performance while limiting the background contributions from the grids. A validation campaign demonstrated tight, sub-mm deflections of the gate and anode, with a combined deflection of 1.8~mm. It showed that the contribution of grid alignment to the detector S2 resolution is expected to be at the \(0.1\%\) level or better. It also informed modifications to the design and production of the extraction region that will reduce the likelihood of breakdown and electron emission. The grids were successfully integrated into the TPC in spring of 2019, and will play an active role in helping LZ achieve a WIMP spin-independent cross-section sensitivity of 1.5\(\times 10^{-48}\) cm\(^{2}\) (90\(\%\) C.L.) at 40 GeV/c\(^{2}\) for the nominal 1,000 day, 5.6 ton fiducial volume exposure~\cite{WIMPsensitivityLZ}.

%% file: 8_Acknowledgements/Acknowledgements.tex
\section{Acknowledgements}

The research supporting this work took place in part at the SLAC National Accelerator Laboratory in Menlo Park, California, the Lawrence Berkeley National Laboratory in Berkeley, California, and the Sanford Underground Research Facility (SURF) in Lead, South Dakota. Funding for this work is supported by the U.S. Department of Energy, Office of Science, Office of High Energy Physics under Contract Numbers DE-AC02-05CH11231, DE-SC0020216, DE-SC0019193, DE-SC0015535, and DE-AC02-76SF00515. This research was also supported by U.S. National Science Foundation (NSF); the University of Wisconsin for grant UW PRJ82AJ; the U.K. Science \& Technology Facilities Council under award numbers ST/M003655/1, ST/S000739/1 and ST/K502042/1 (AB). This research used resources of the National Energy Research Scientific Computing Center, a DOE Office of Science User Facility supported by the Office of Science of the U.S. Department of Energy under Contract No. DE-AC02-05CH11231. We acknowledge many types of support provided to us by the South Dakota Science and Technology Authority (SDSTA), which developed the Sanford Underground Research Facility (SURF) with an important philanthropic donation from T. Denny Sanford as well as support from the State of South Dakota. SURF is operated by the SDSTA under contract to the Fermi National Accelerator Laboratory for the DOE, Office of Science. 

We also gratefully acknowledge the technical support and the access to laboratory space, scientists and technicians, and resources made available by the SLAC National Accelerator Laboratory, the Lawrence Berkeley National Laboratory, and the Sanford Underground Research Facility. 

We also wish to thank a number of colleagues who participated in the weaving and fabrication of the prototype and final LZ grids: Liz Atkin, Jonathan Daniel, Rayan Sud, Labib Tazwar Rahman, Margaret Koulikova, Madan Timalsina, Alexander Madurowicz, Jacob Cutter, Andreas Biekert, and Julia Baumgarte.

We are also grateful for to our collaborators Peter Sorensen at Lawrence Berkeley National Laboratory, Jingke Xu at Lawrence Livermore National Laboratory, and Kelsey Oliver-Mallory at Imperial College London for their careful revision of and comments on this manuscript.

We also acknowledge the support of Astro Pak Corp. for accommodating our efforts to clean our various grid rings and passivate the gate. We also thank Advanced Electropolishing for likewise working with us throughout the grid R\(\&\)D and production process.

%% file: main.bbl
\begin{thebibliography}{10}
\expandafter\ifx\csname url\endcsname\relax
  \def\url#1{\texttt{#1}}\fi
\expandafter\ifx\csname urlprefix\endcsname\relax\def\urlprefix{URL }\fi
\expandafter\ifx\csname href\endcsname\relax
  \def\href#1#2{#2} \def\path#1{#1}\fi

\bibitem{BACCARAT}
D.~S. Akerib, et~al., {Simulations of events for the LUX-ZEPLIN (LZ) dark
  matter experiment}, Astroparticle Physics 125 (2021) 102480.
\newblock \href
  {http://dx.doi.org/https://doi.org/10.1016/j.astropartphys.2020.102480}
  {\path{doi:https://doi.org/10.1016/j.astropartphys.2020.102480}}.

\bibitem{LZDetector}
D.~S. Akerib, et~al., {The LUX-ZEPLIN (LZ) experiment}, Nuclear Instruments and
  Methods in Physics Research Section A: Accelerators, Spectrometers, Detectors
  and Associated Equipment 953 (2020) 163047.
\newblock \href {http://dx.doi.org/https://doi.org/10.1016/j.nima.2019.163047}
  {\path{doi:https://doi.org/10.1016/j.nima.2019.163047}}.

\bibitem{rpp2019}
{M. Tanabashi et al.}, {(Particle Data Group)}, {Phys. Rev. D 98, 030001,
  (2018)}.

\bibitem{B8}
R.~Essig, M.~Sholapurkar, T.-T. Yu, Solar neutrinos as a signal and background
  in direct-detection experiments searching for sub-gev dark matter with
  electron recoils, Phys. Rev. D 97 (2018) 095029.
\newblock \href {http://dx.doi.org/10.1103/PhysRevD.97.095029}
  {\path{doi:10.1103/PhysRevD.97.095029}}.

\bibitem{DoubleBetaDecaySensitivity}
D.~S. Akerib, et~al., {Projected sensitivity of the LUX-ZEPLIN experiment to
  the $0\ensuremath{\nu}\ensuremath{\beta}\ensuremath{\beta}$ decay of
  $^{136}\mathrm{Xe}$}, Phys. Rev. C 102 (2020) 014602.
\newblock \href {http://dx.doi.org/10.1103/PhysRevC.102.014602}
  {\path{doi:10.1103/PhysRevC.102.014602}}.

\bibitem{LUXDiscrimination}
D.~S. Akerib, et~al., {Discrimination of electronic recoils from nuclear
  recoils in two-phase xenon time projection chambers}, Phys. Rev. D 102 (2020)
  112002.
\newblock \href {http://dx.doi.org/10.1103/PhysRevD.102.112002}
  {\path{doi:10.1103/PhysRevD.102.112002}}.

\bibitem{EXODriftVelocity}
J.~B. Albert, et~al., {Measurement of the drift velocity and transverse
  diffusion of electrons in liquid xenon with the EXO-200 detector}, Phys. Rev.
  C 95 (2017) 025502.
\newblock \href {http://dx.doi.org/10.1103/PhysRevC.95.025502}
  {\path{doi:10.1103/PhysRevC.95.025502}}.

\bibitem{LivermoreExtractionEfficiency}
J.~Xu, et~al., Electron extraction efficiency study for dual-phase xenon dark
  matter experiments, Phys. Rev. D 99 (2019) 103024.
\newblock \href {http://dx.doi.org/10.1103/PhysRevD.99.103024}
  {\path{doi:10.1103/PhysRevD.99.103024}}.

\bibitem{LUXEBackgrounds}
D.~S. Akerib, et~al., {Investigation of background electron emission in the LUX
  detector}, Phys. Rev. D 102 (2020) 092004.
\newblock \href {http://dx.doi.org/10.1103/PhysRevD.102.092004}
  {\path{doi:10.1103/PhysRevD.102.092004}}.

\bibitem{ChepelAraujo}
V.~Chepel, H.~Ara{\'u}jo, Liquid noble gas detectors for low energy particle
  physics, Journal of Instrumentation 8~(04) (2013) R04001.

\bibitem{ManninoThesis}
R.~L. Mannino, {Measuring Backgrounds From \(^{85}\)Kr and \(^{210}\)Bi to
  Improve Sensitivity of Dark Matter Detectors}, Ph.D. thesis, Texas A\&M
  University (Jan 2017).

\bibitem{X1TEnergyResolution}
E.~Aprile, et~al., {Energy resolution and linearity of XENON1T in the MeV
  energy range}, The European Physical Journal C 80~(8) (2020) 1--9.

\bibitem{BaileyThesis}
A.~Bailey, {Dark Matter Searches and Study of Electrode Design in LUX and LZ.
  }, Ph.D. thesis, Imperial College London (2016).
\newblock \href {http://dx.doi.org/10.25560/41878} {\path{doi:10.25560/41878}}.

\bibitem{ImperialWireStudies}
A.~Tom\'{a}s, et~al., Study and mitigation of spurious electron emission from
  cathodic wires in noble liquid time projection chambers, Astroparticle
  Physics 103 (2018) 49--61.
\newblock \href
  {http://dx.doi.org/https://doi.org/10.1016/j.astropartphys.2018.07.001}
  {\path{doi:https://doi.org/10.1016/j.astropartphys.2018.07.001}}.

\bibitem{FNALHVWorkshop}
B.~Rebel, et~al., {High voltage in noble liquids for high energy physics},
  Journal of Instrumentation 9~(08) (2014) T08004.

\bibitem{XENON100}
E.~Aprile, et~al., {The XENON100 dark matter experiment}, Astroparticle Physics
  35~(9) (2012) 573--590.

\bibitem{ZEPLIN-IIIDetector}
D.~Y. Akimov, et~al., {The ZEPLIN-III dark matter detector: Instrument design,
  manufacture and commissioning}, Astroparticle Physics 27~(1) (2007) 46--60.

\bibitem{asm}
{MatWeb Material Property Data}, {304 Stainless Steel.},
  \url{http://www.matweb.com/search/datasheet.aspx?MatGUID=abc4415b0f8b490387e3c922237098da}.

\bibitem{engineeringtoolbox}
{The Engineering Toolbox}, {Coefficients of Linear Thermal Expansion},
  \url{https://www.engineeringtoolbox.com/linear-expansion-coefficients-d_95.html}.

\bibitem{cfw}
{California Fine Wire}, {Stainless Steel 304 Material \(\#\): 100192},
  \url{https://calfinewire.com/datasheets/100192-stainlesssteel304/}.

\bibitem{copperReflectivity}
E.~O. Hulburt, {The Reflecting Power of Metals in the Ultra-Violet Region of
  the Spectrum}, APJ 42 (1915) 205.
\newblock \href {http://dx.doi.org/10.1086/142202} {\path{doi:10.1086/142202}}.

\bibitem{goldReflectivity}
G.~Shuyi, et~al., Influence of binding layer on the reflective performance of a
  au film in vacuum ultraviolet wavelength region, Applied optics 46 (2008)
  8641--4.
\newblock \href {http://dx.doi.org/10.1364/AO.46.008641}
  {\path{doi:10.1364/AO.46.008641}}.

\bibitem{alReflectivity}
S.~Wilbrandt, et~al., {Protected and enhanced aluminum mirrors for the VUV},
  Appl. Opt. 53~(4) (2014) A125--A130.
\newblock \href {http://dx.doi.org/10.1364/AO.53.00A125}
  {\path{doi:10.1364/AO.53.00A125}}.

\bibitem{ssReflectivity}
S.~Bricola, et~al., {Noble-gas liquid detectors: measurement of light diffusion
  and reflectivity on commonly adopted inner surface materials}, Nuclear
  Physics B - Proceedings Supplements 172 (2007) 260--262, proceedings of the
  10th Topical Seminar on Innovative Particle and Radiation Detectors.
\newblock \href
  {http://dx.doi.org/https://doi.org/10.1016/j.nuclphysbps.2007.08.059}
  {\path{doi:https://doi.org/10.1016/j.nuclphysbps.2007.08.059}}.

\bibitem{conceptualDesignReportLZ}
D.~S. Akerib, et~al., {LUX-ZEPLIN (LZ) Conceptual Design Report}, Tech. Rep.
  LBNL-190005, FERMILAB-TM-2621-AE-E-PPD (September 2015).
\newblock \href {http://arxiv.org/abs/1509.02910} {\path{arXiv:1509.02910}}.

\bibitem{fieldEmissionTheory}
R.~G. Forbes, in: {2016 Young Researchers in Vacuum Micro/Nano Electronics
  (VMNE-YR)}, title={Field electron emission theory (October 2016)}, 2016, pp.
  1--8.
\newblock \href {http://dx.doi.org/10.1109/VMNEYR.2016.7880403}
  {\path{doi:10.1109/VMNEYR.2016.7880403}}.

\bibitem{Williams_1974}
D.~W. Williams, W.~T. Williams, Prebreakdown and breakdown characteristics of
  stainless steel electrodes in vacuum, Journal of Physics D: Applied Physics
  7~(8) (1974) 1173--1183.
\newblock \href {http://dx.doi.org/10.1088/0022-3727/7/8/315}
  {\path{doi:10.1088/0022-3727/7/8/315}}.

\bibitem{plasmaChemistry}
A.~Fridman, {Plasma Chemistry}, Cambridge University Press, 2008.

\bibitem{garfieldpp}
{The Garfield++ Collaboration},
  \url{http://garfieldpp.web.cern.ch/garfieldpp/}.

\bibitem{tdr}
B.~J. Mount, et~al., {LUX-ZEPLIN (LZ) Technical Design Report}, Tech. Rep.
  LBNL-1007256, FERMILAB-TM-2653-AE-E-PPD (March 2017).
\newblock \href {http://arxiv.org/abs/1703.09144} {\path{arXiv:1703.09144}}.

\bibitem{Boyle}
G.~J. Boyle, et~al., Ab initio electron scattering cross-sections and transport
  in liquid xenon, Journal of Physics D: Applied Physics 49~(35) (2016) 355201.

\bibitem{NEST}
M.~Szydagis, et~al., {Noble element simulation technique v2.0.1, (2019)}.
\newblock \href {http://dx.doi.org/10.5281/zenodo.3357973}
  {\path{doi:10.5281/zenodo.3357973}}.

\bibitem{historicalLooms}
T.~Kuroda, H.~Takahashi, A.~Masuda, {Wearable Sensors}, Academic Press, Oxford,
  2014.

\bibitem{betaCage}
R.~Schnee, et~al., {Removal of long-lived \(^{222}\)Rn daughters by
  electropolishing thin layers of stainless steel}, AIP Conference Proceedings
  1549~(1) (2013) 128--131.
\newblock \href {http://dx.doi.org/10.1063/1.4818092}
  {\path{doi:10.1063/1.4818092}}.

\bibitem{SLACcitricpaper}
D.~S. Akerib, et~al., {Reduction of Electron Emission Rate from Grid Electrodes
  with Single Electron Sensitivity}, to appear.

\bibitem{OxideNoer}
R.~Noer, Electron field emission from broad-area electrodes, Applied Physics A
  28~(1) (1982) 1--24.

\bibitem{OxideStygar}
W.~A. Stygar, et~al., Suppression of electron emission from metal electrodes:
  Ldrd 28771 final report., Tech. rep., Sandia National Laboratories (2003).

\bibitem{OxideHuang}
Q.-A. Huang, Field emission from silicon covered with a thin oxide layer, in:
  IVMC'95. Eighth International Vacuum Microelectronics Conference. Technical
  Digest (Cat. No. TH8012), IEEE, 1995, pp. 373--377.

\bibitem{OxideYang}
G.~Yang, K.~K. Chin, R.~B. Marcus, Electron field emission through a very thin
  oxide layer, IEEE Transactions on Electron Devices 38~(10) (1991) 2373--2376.

\bibitem{kitchenSinkST}
D.~S. Akerib, et~al., {A platform for the development of large-scale
  circulation, cryogenics and high voltage systems for the LZ experiment}To
  appear.

\bibitem{cleanlinessPaper}
D.~S. Akerib, et~al., {The LUX-ZEPLIN (LZ) radioactivity and cleanliness
  control programs}, The European Physical Journal C 80~(11) (2020) 1--52.

\bibitem{SorensenLBL}
P.~Sorensen, Private Communication.

\bibitem{WIMPsensitivityLZ}
{D. S. Akerib et al.}, {Projected WIMP sensitivity of the LUX-ZEPLIN dark
  matter experiment}, {Phys. Rev. D} {101} ({2020}) {052002}.
\newblock \href {http://dx.doi.org/{10.1103/PhysRevD.101.052002}}
  {\path{doi:{10.1103/PhysRevD.101.052002}}}.

\end{thebibliography}
